\documentclass[fleqn,usenatbib]{mnras}

\usepackage{graphicx}	% Including figure files
\usepackage{amsmath}	% Advanced maths commands
\usepackage{amssymb}	% Extra maths symbols
\usepackage{multirow}
\usepackage{multicol}        % Multi-column entries in tables
\usepackage{bm}		% Bold maths symbols, including upright Greek
\usepackage{pdflscape}	% Landscape pages
\usepackage{lineno}
\usepackage[dvipsnames]{xcolor}
\usepackage[T1]{fontenc}
\usepackage{ae,aecompl}

\usepackage{newtxtext,newtxmath}
\title[Gamma-Ray Polarimetry of the Crab Pulsar Observed by POLAR]{\textit{Gamma-Ray Polarimetry of the Crab Pulsar Observed by POLAR}}

\author[H. C. Li et al.]{Han-Cheng Li$^{1,2,3}$\thanks{Contact e-mail: \href{mailto:hancheng.li@unige.ch}{hancheng.li@unige.ch}},
Nicolas Produit$^{1}$,
Shuang-Nan Zhang$^{2,3}$,
Merlin Kole$^{4}$,
Jian-Chao Sun$^{2,3}$,
\newauthor
Ming-Yu Ge$^{2,3}$,
Nicolas De Angelis$^{4}$,
Johannes Hulsman$^{4}$,
Zheng-Heng Li$^{5}$,
Li-Ming Song$^{2,3}$,
\newauthor
Teresa Tymieniecka$^{6}$,
Bo-Bing Wu$^{2}$,
Xin Wu$^{4}$,
Yuan-Hao Wang$^{2}$,
Shao-Lin Xiong$^{2}$,
Yong-Jie Zhang$^{2}$,
\newauthor
Yi Zhao$^{2,7}$,
and Shi-Jie Zheng$^{2}$
\\
$^{1}$Department of Astronomy, University of Geneva, 16 Chemin d'Ecogia, 1290 Versoix, Switzerland \\
$^{2}$Key Laboratory of Particle Astrophysics, Institute of High Energy Physics, Chinese Academy of Sciences, Beijing 100049, China \\ 
$^{3}$University of Chinese Academy of Sciences, Beijing 100049, China \\
$^{4}$Department of Nuclear and Particle Physics, University of Geneva, 24 Quai Ernest-Ansermet, 1205 Geneva, Switzerland \\
$^{5}$Center for Transformative Science, ShanghaiTech University, Shanghai 201210, China \\
$^{6}$National Centre for Nuclear Research, ul. A. Soltana 7, 05-400 Otwock, Swierk, Poland \\
$^{7}$Department of Astronomy, Beijing Normal University, Beijing 100875, China
}

\date{22 February 2022}

\pubyear{2022}

\begin{document}
\label{firstpage}
\pagerange{\pageref{firstpage}--\pageref{lastpage}}
\maketitle

\begin{abstract}
The X/$\gamma$ ray polarimetry of the Crab pulsar/nebula is believed to hold crucial information on their emission models. In the past, several missions have shown evidence of polarized emission from the Crab. The significance of these measurements remains however limited. New measurements are therefore required. POLAR is a wide Field of View Compton-scattering polarimeter (sensitive in 50-500 keV) onboard the Chinese spacelab Tiangong-2 which took data from September 2016 to April 2017. Although not designed to perform polarization measurements of pulsars, we present here a novel method which can be applied to POLAR as well as that of other wide Field of View polarimeters. The novel polarimetric joint-fitting method for the Crab pulsar observations with POLAR, allows us to obtain constraining measurements of the pulsar component.  The best fitted values and corresponding 1$\sigma$ deviations for the averaged phase interval: (PD=$14\substack{+15 \\ -10}$\%, PA=$108\substack{+33 \\ -54} ^{\circ}$), for Peak 1: (PD=$17\substack{+18 \\ -12}$\%, PA=$174\substack{+39 \\ -36} ^{\circ}$) and for Peak 2: (PD=$16\substack{+16 \\ -11}$\%, PA=$78\substack{+39 \\ -30} ^{\circ}$). Further more, the 3$\sigma$ upper limits on the polarization degree are for the averaged phase interval (55\%), Peak 1 (66\%) and Peak 2 (57\%). Finally, to illustrate the capabilities of this method in the future, we simulated two years observation to the Crab pulsar with POLAR-2. The results show that POLAR-2 is able to confirm the emission to be polarized with $5\sigma$ and $4\sigma$ confidence level if the Crab pulsar is polarized at $20\%$ and $10\%$ respectively.
\end{abstract}

\begin{keywords}
Instrumentation: polarimeters 
-- Methods: data analysis
-- Stars: pulsars: individual
-- Techniques: polarimetric
-- Gamma rays: general
\end{keywords}

%%%%%%%%%%%%%%%%%%%%%%%%%%%%%%%%%%%%%%%%%%%%%%%%%%

%%%%%%%%%%%%%%%%% BODY OF PAPER %%%%%%%%%%%%%%%%%%

% The MNRAS class isn't designed to include a table of contents, but for this document one is useful.
% I therefore have to do some kludging to make it work without masses of blank space.

%\begingroup
%\let\clearpage\relax
%\tableofcontents
%\endgroup
%\newpage
%\linenumbers

\section{Introduction}\label{sec:introduction}

The Crab pulsar (PSR B0531+21) is located in the heart of the Crab nebula (supernova remnant of SN 1054). As an isolated pulsar, the Crab pulsar is rotation-powered with $\dot{E}\approx5\times{10^{38}}$ erg/s. Electromagnetic radiation is emitted from its co-rotating strong dipole magnetic field. The Crab pulsar shows a double-pulse structure over all wavelengths, with pulse phase intervals defined as Peak 1 (P1), Peak 2 (P2), and the Off Pulse (OP) where the nebula dominates. 
The broadband high brightness makes it widely studied over the past five decades by monitoring its spectral, timing and imaging properties. Thereby multiple emission models have been developed, including polar gap model~\citep{1975ApJ...196...51R}, outer gap model~\citep{1986ApJ...300..500C, 1986ApJ...300..522C}, slot gap model~\citep{2008ApJ...688L..25H} and annular gap model~\citep{2010MNRAS.406.2671D}, etc.
Despite all these efforts, detailed physical questions related to the emission mechanism and region, geometry of source, etc., are yet to be precisely understood. Polarimetry is widely believed to be a unique probe to answer some of these questions; phase resolved polarimetry is more crucial since the observed emission regions are phase-dependent. Some theoretical predictions have been made by \cite{2007ApJ...656.1044T, 2013MNRAS.434.2636P, 2017ApJ...840...73H}. The two parameters of polarimetry are commonly referred as Polarization Degree (PD) and Polarization Angle (PA). Although the actual polarization of sources can be a mixture of linear and circular components, the measurements of current X/$\gamma$ ray polarimeters are limited to the linear polarization only. Therefore, in this and all cited works here, we only refer to the linear polarization.

The OSO-8 mission with Bragg crystal polarimeters performed the first successful astrophysical polarization measurements in the X-ray energy band. These measurements were of the Crab nebula at energies of 2.6 keV  and 5.2 keV  \citep{1976ApJ...208L.125W,1978ApJ...220L.117W}. At 2.6 keV the results are (PD=15.7$\pm$1.5\%, PA=161.1$\pm$2.8$^{\circ}$) for Averaged Phase interval of the Crab (AP, including pulsar and nebula), (PD=19.2$\pm$1.0\%, PA=156.4$\pm$1.4$^{\circ}$) for OP (only nebula), and similar results were found at $\sim$ 5.2 keV. These results support the theory that synchrotron radiation is responsible for the X-ray emission from the Crab nebula. 
Long after that, the SPI and IBIS on board the INTEGRAL mission measured the hard X-ray polarization measurements for the Crab nebula and pulsar by accumulating Compton-scattering events. The main results can be summarized as \cite{2008ApJ...688L..29F} (200-800 keV): AP (PD=$47\substack{+19 \\ -13}$\%, PA=100$\pm$11$^{\circ}$), P1+P2 (PD=$42\substack{+30 \\ -16}$\%, PA=70$\pm$20$^{\circ}$), OP (PD>72\%, PA=120.6$\pm$8.5$^{\circ}$); \cite{2008Sci...321.1183D} (100-1000 keV): OP (PD=47$\pm$10\%, PA=123$\pm$11$^{\circ}$); \cite{2013ApJ...769..137C} (130-400 keV): AP (PD=28$\pm$6\%, PA=117$\pm$9$^{\circ}$); \cite{2019ApJ...882..129J} (130-400 keV): AP (PD=24$\pm$4\%, PA=120$\pm$6$^{\circ}$). However, the first two publications indicated a high PD for the Crab, whereas the latter two obtained much lower PD measurements. Additionally, as neither of the two INTEGRAL instruments is a dedicated polarimeter, no on-ground calibration was performed. As a result, one cannot validate the analysis and the magnitude of the systematic uncertainty on the measurements.

The polarization parameters of the Crab pulsar and nebula have been precisely measured at optical wavelengths. The latest optical polarization measurements on the Crab pulsar and nebula were done by
OPTIMA (610 and 1400 MHz) \citep{2009MNRAS.397..103S} and HST \citep{2013MNRAS.433.2564M}. For the Crab pulsar: 1) OPTIMA measured for the AP (PD=5.5$\pm$0.1\%, PA=96.4$\pm$0.1$^{\circ}$), and the phase-resolved results show that PDs are higher in the bridge and off-pulse region than those in the two peaks, and the PAs swing through a large angle in both peaks; 2) HST measured for the AP (PD=5.2$\pm$0.3\%, PA=105.1$\pm$1.6$^{\circ}$).  For the Crab nebula: OPTIMA measured (PD=9.7$\pm$0.1\%, PA=139.8$\pm$0.2$^{\circ}$) when the region is near the pulsar; the averaged PA value is 140$^{\circ}$ in the region 2-3 arcsec away from the pulsar; beyond 5 arcsec the PA exceeds 155$^{\circ}$ and becomes highly position dependent. 
	
In recent years, several dedicated Compton-scattering or Photoelectron-track polarimeters have obtained more precise polarization measurements in X/$\gamma$ ray bands \citep{2021JApA...42..106C}. For example, PoGO+ (Compton, 20-160 keV, \cite{2017NatSR...7.7816C}) measured polarization for AP (PD=20.9$\pm$5.0\%, PA=131.3$\pm$6.8$^{\circ}$), P1 (PD=$0\substack{+29 \\ -0}$\%, PA=153$\pm$43$^{\circ}$), P2 (PD=$33.5\substack{+18.6 \\ -22.3}$\%, PA=153$\pm$43$^{\circ}$) and OP (PD=$17.4\substack{+8.6 \\ -9.3}$\%, PA=137$\pm$15$^{\circ}$). Their results also rejected a high PD. AstroSat CZTI (Compton, 100-380 keV, \cite{2018NatAs...2...50V}) measured polarization for AP (PD=32.7$\pm$5.8\%, PA=143.5$\pm$2.8$^{\circ}$) and OP (39.0$\pm$10.0\%, 140.9$\pm$3.7$^{\circ}$). Based on further phase-resolved results, they claimed that there are variable polarization properties within the off-pulse region and a swing of the polarization angle across the pulse peaks. Hitomi SGD (Compton, 60-160 keV, \cite{2018PASJ...70..113H}) also obtained a significant polarization measurement for AP of (PD=22.1$\pm$10.6\%, PA=$110.7\substack{+13.2 \\ -13.0} ^{\circ}$) with 5 ks short exposure time. More recently, a collimated CubeSate PolarLight (Photoelectron, 3.0–4.5 keV, \cite{2020NatAs...4..511F}) measured polarization for AP (PD=$15.3\substack{+3.1 \\ -3.0}$\%, PA=145.8$\pm$5.7$^{\circ}$), P1+P2 (PD=15.8$\pm$3.9\%, PA=147.6$\pm$7.0$^{\circ}$), OP (PD=$14.0\substack{+5.2 \\ -5.4}$\%, PA=142.4$\pm$11.0$^{\circ}$). They further performed time-dependent polarization analysis, and found that PD decreased after the glitch of the Crab pulsar on 23 July 2019. The polarization results in X/$\gamma$ ray bands mentioned above are listed in Table~\ref{table:PDPA_resultstable}.

POLAR is a dedicated Gamma-Ray Burst (GRB) Compton-scattering polarimeter sensitive in 50-500 keV energy range \citep{2005NIMPA.550..616P, 2018NIMPA.877..259P}. POLAR was launched on September 2016 onboard the second Chinese space laboratory Tiangong-2 (TG-2). The zenith of POLAR was continuously pointing to the sky, and it was working in a scanning observation mode while running it during its orbit. Given the random occurrence of GRBs in the sky, POLAR is equipped with a wide Field of View (FoV) ($\sim2\pi$ Sr) and a large effective area (400 cm$^{2}$ at 300 keV). These features make it one of the most sensitive instruments for GRB detections in its energy range. During its six-month observation, POLAR detected 55 GRBs \citep{2017ICRC...35..640X}. In order to estimate systematic effects, POLAR has been carefully calibrated both on-ground and in-orbit \citep{2017NIMPA.872...28K, 2018NIMPA.900....8L} for polarization and spectral responses. Thereafter, 14 of the detected GRBs have been used to preform polarization analyses \citep{2019NatAs...3..258Z, 2019A&A...627A.105B, 2020A&A...644A.124K}.
	
Apart from transients, such as GRBs, POLAR cannot observe persistent celestial sources due to their low Signal to Noise Ratio (SNR) resulted from POLAR's large FoV. However, POLAR can monitor some luminous X-ray pulsars. The periodic nature of pulsars makes it possible to stack the signal by aligning pulse phase values. Thanks to the well established timing analysis for pulsars, the pulse phase parameter can be derived. Based on the pulse phase parameter, spectrum and modulation curves can be additionally extracted for further studies. POLAR has detected folded pulsations from both the Crab pulsar and the PSR B1509-58 significantly (detailed analyses on observation and simulation verification can be found in \cite{2017ICRC...35..820L, 2021arXiv210903142L}). Since POLAR is not dedicated for pulsar studies, the SNR of pulsar detection is much lower than that of GRBs, but this can partially be compensated using the larger exposure time for pulsars. Limited by the low SNR in the end, PSR B1509-58 has not yet been further studied. While the Crab pulsar detection was much more significant, and was firstly chosen to test pulsar navigation algorithm to predict the TG-2 orbital parameters \citep{2017SSPMA..47i9505Z}. As POLAR is a non-pointing wide FoV instrument, the observation datasets are complicated by multiple observational parameters like incoming photon angles. Therefore, before proceeding with the polarization studies of the Crab pulsar using POLAR data, we performed a phase-resolved spectroscopy for checking the spectral responses from all incoming angles \citep{2019JHEAp..24...15L}. Previous timing and spectral studies of the Crab pulsar with POLAR data served as timing and spectral calibrations to the POLAR instrument, and lay the groundwork for the polarization study.
	
In this paper, we present the polarimetric methodology developed as well as the resulting polarization measurements of the Crab pulsar with POLAR. In Section~\ref{sec:observation} we introduce the observations and data selection criteria. The methodology of the polarimetry of the Crab pulsar is described in Section~\ref{sec:methodology}. In Section~\ref{sec:result}, phase-resolved polarization results are presented. Subsequently, in Section~\ref{sec:polar-2}, we show the future mission POLAR-2's prospective polarimetry on the Crab pulsar. Finally, we draw the main conclusions of this paper in Section~\ref{sec:summary}.
	
%-------------------------------------------------------------------------------------------
	
\section{Observation}\label{sec:observation}

POLAR took approximately six months in-orbit observation data from September 2016 to April 2017, during which the Crab pulsar was visible to POLAR when it passed through POLAR's FoV. The timing analysis used here for the Crab pulsar is based on previous work \citep{2017ICRC...35..820L, 2019JHEAp..24...15L, 2021arXiv210903142L}. Note that phase zero reference in our analysis is taken from the X-ray observation. The general data pipeline is based on the standard method described in ~\citep{2018NIMPA.900....8L}. This pipeline handles processes such as baseline/noise subtraction, signal crosstalk correction, energy reconstruction, as well as filtering of cosmic-ray events and post-cosmic-ray events. Additionally, the Good Time Interval (GTI) were generated by applying a count rate threshold on the one-second-binning light curves. With GTI we mainly excluded high background intervals caused by the South Atlantic Anomaly (SAA), the high latitude areas, GRBs, solar flares, etc. With GTI selection we obtain a total of $\sim$4400 ks exposure time to the Crab pulsar.

\begin{figure}
	\centering
	\includegraphics[width=8cm,height=6cm]{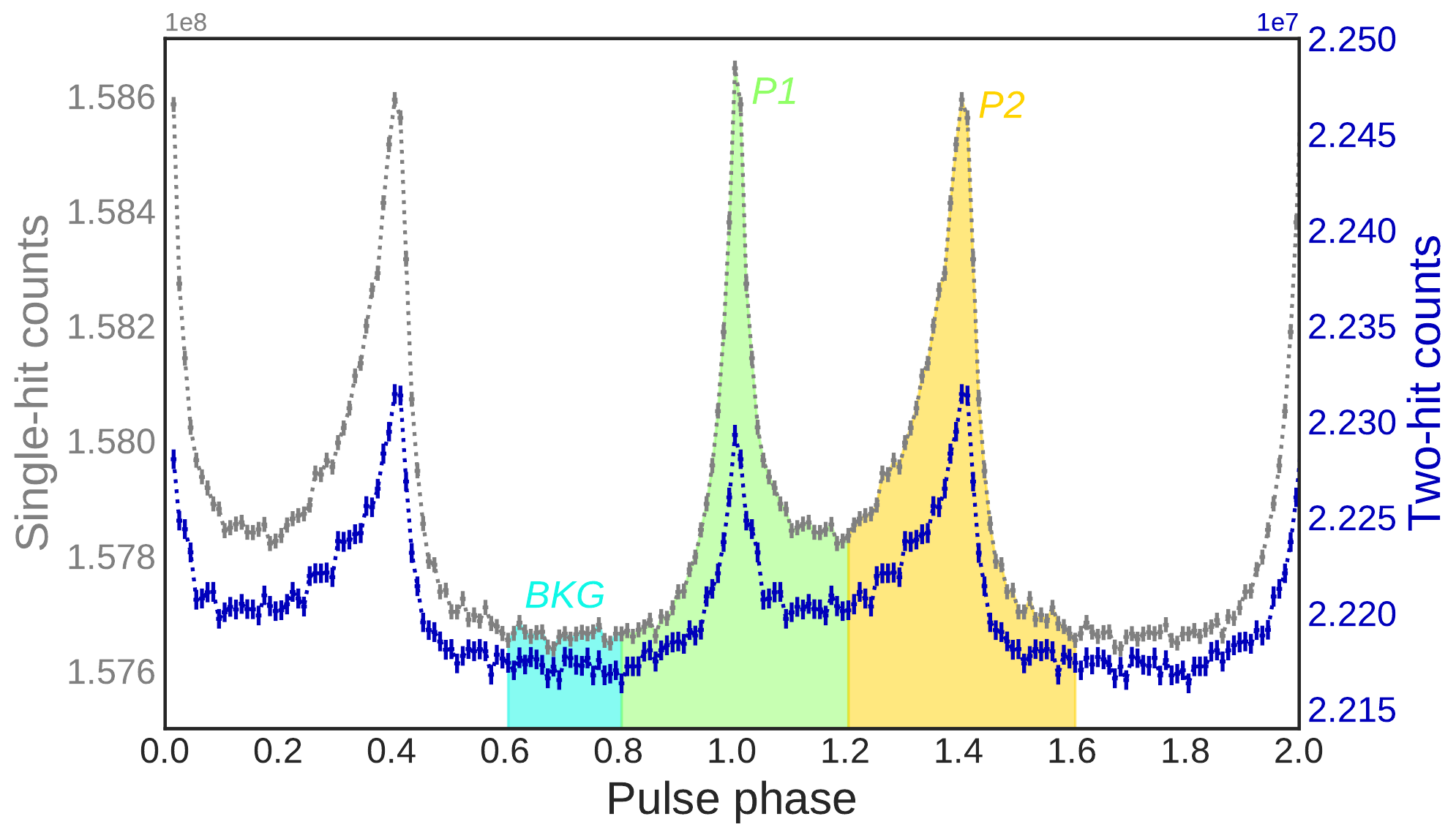}
	\caption{\textbf{Pulse profiles of single-hit and two-hit events.} The definitions of different phase intervals in this work are defined as: OP (0.6-0.8), P1 (0-0.2 || 0.8-1.0), P2 (0.2-0.6). OP contains the nebula and background photons that cannot be distinguished by POLAR, so the whole OP is treated as background (BKG). The colored regions on the pulse profiles display the different phase intervals (shifted 1 phase to make it clear). Note that only two-hit events (which contain Compton-scattering events) are used for the polarization analysis.}
	\label{fig:obs_vs_sd}
\end{figure}

Using this selection the pulse profile can be folded versus different observational parameters. Figure \ref{fig:obs_vs_sd} shows the pulse profiles of the so-called single-hit events and two-hit events. The two-hit events contain at least two channels within POLAR above the trigger threshold while the single-hit events only contain one triggered channel. This distinction is important as only two-hit events can be used for polarization studies while the single-hit events can only be used for timing and spectral analysis such as in \citep{2019JHEAp..24...15L}. The two profiles are precisely aligned, but two-hit counts are $\sim$14\% of the single-hit counts. The definitions of the different phase intervals in this work are: P1 (0-0.2 || 0.8-1.0), P2 (0.2-0.6) and OP (0.6-0.8). As a wide FoV detector, POLAR cannot distinguish pulsar, nebula and background (signal from neither pulsar nor nebula ) components spatially, so the pulse phase is the only parameter that can be used for dividing these components. We assume that the nebula and background are phase independent, and that the pulsar has no contribution in the OP interval. In this case, the OP interval contains only the nebula and background components, which can be subtracted from P1 and P2 intervals to get pure pulsar emissions. To summarize, we cannot extract the nebula component from the  background, so that both of them are subtracted from the on-pulse intervals. Therefore, the polarization measurements in this work are only for pulsar emissions without the nebula contribution.

\begin{figure}
	\centering
	\includegraphics[width = 7cm, height = 7cm]{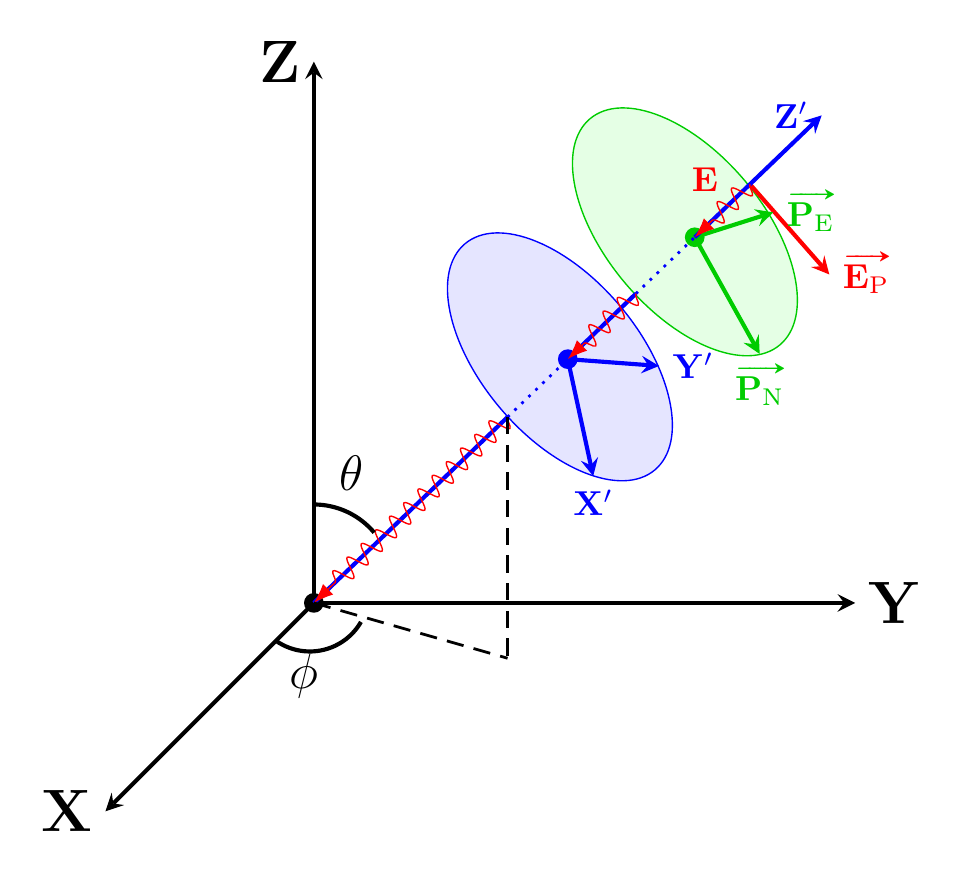}
	\caption{\textbf{Coordinate definitions in this work.} The POLAR coordinate system (in black) is the local coordinate system for the POLAR detector, when photons (in energy E) of a source arrive at POLAR, the incident angle of them can be described as $\theta$ (angle from POLAR's zenith angle) and $\phi$ (azimuthal angle). The source coordinate (in blue) is rotated from the POLAR coordinate by $\theta$ and $\phi$, its Z-axis points from the detector to the source (inverse to the source's incident direction), and the X$'$-Y$'$ plane is perpendicular to the source's incident direction. The J2000/Equatorial coordinate (in green) is the standard source location coordinate system, the North-East (projections onto the source plane are \textbf{P$_{\textbf{N}}$}-\textbf{P$_{\textbf{E}}$}) plane is used to define the PA value (from the North to the East) and to demonstrate the polarization vector \textbf{E$_{\textbf{P}}$}. Detailed projection illustration is shown in Figure \ref{fig:projection_plane}.}\label{fig:coordinates}
\end{figure}

Before investigating other observational parameters, we have to introduce three different coordinate systems used in this work as shown in Figure \ref{fig:coordinates}, which are: 1) POLAR coordinates, the local coordinates of POLAR detector, the incoming angle of a source can be described as $\theta$ (zenith angle) and $\phi$ (azimuthal angle); we note that the definitions of $\theta$ and $\phi$ here are different from the ones in Figure \ref{fig:scattering}; 2) source coordinates, its Z-axis points from the detector to the source (inverse to the source's incident direction), and the X$'$-Y$'$ plane is perpendicular to the source's incident direction; 3) J2000/Equatorial coordinates, the North-East plane of which is commonly used to define the PA value (from the North to the East) and to demonstrate the polarization vector. First, unlike the GRB observation, which happens within a short duration where the incident angle of the source in the POLAR local coordinate does not change significantly, the Crab pulsar observation angle relative to the detector varies over the entire mission so that data with multiple incident angles need to be separated. In the previous work of \cite{2017ICRC...35..820L}, the pulse profile versus the incoming angle $\theta$ shows that the Crab pulsar is visible to POLAR when $\theta$ is below $\sim102^{\circ}$. In order to resolve this parameter, we employed HealPix (HP, \cite{2005ApJ...622..759G}) to divide POLAR's local sky into 768 HP bins (numbering is 0-767), where $N_{\rm side}=8$ and RING numbering schedule is used. Given the conditional visibility, the first 400 HP bins (0-399) are considered, more details can be found in \citep{2019JHEAp..24...15L}.
	
\begin{figure}
	\centering
	\includegraphics[width=8cm,height=6cm]{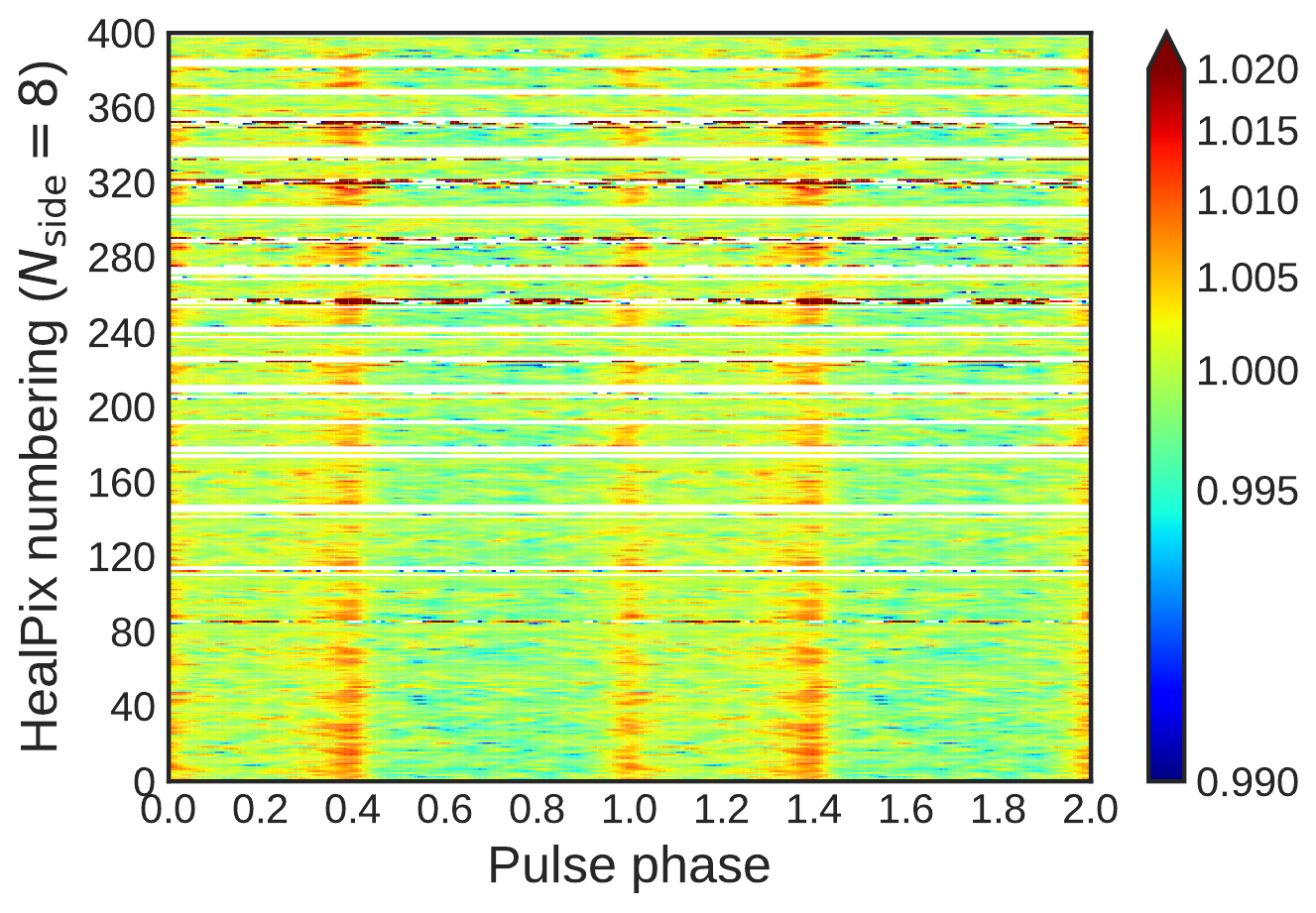}	
	\caption{\textbf{Pulse profiles of two-hit events v.s the healpix bin (HP).} The z-axis contains the normalized counts of the pulse profile of each HP bin. The statistics of signal vary with the HP bins since the exposure time and sensitivity of each HP bin is different.}
	\label{fig:obs_vs_hp}
\end{figure}

\begin{figure}
	\centering
	\includegraphics[width=7cm,height=6cm]{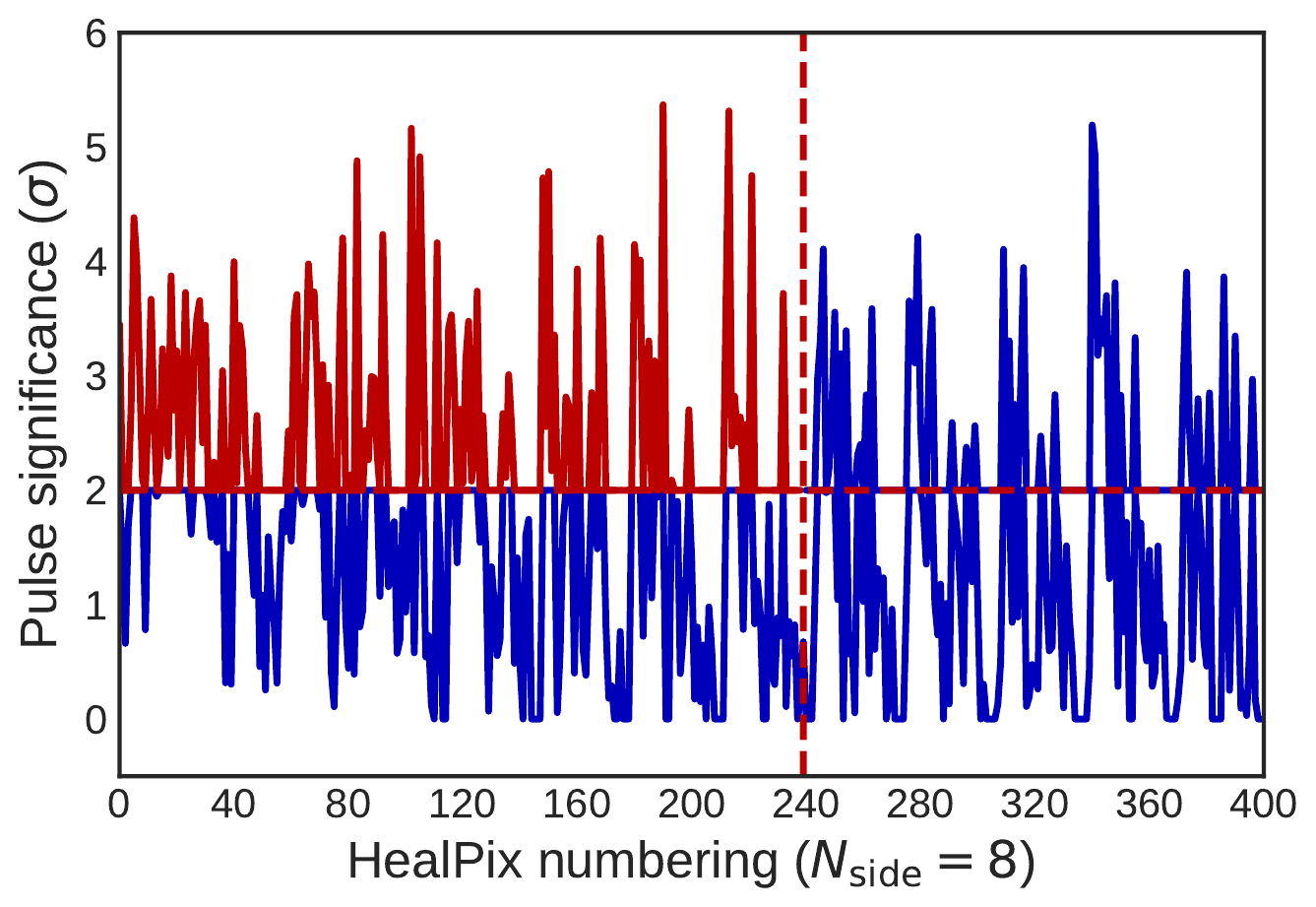}
	\caption{\textbf{Pulse significance of different healpix bins.}  The horizontal red dashed line (2 $\sigma$) shows the applied cut on pulse significance, while the vertical one (239 HP number) shows the applied cut on HP bins which is introduced in Appendix \ref{app:systematic}. 108 HP bins (red curves) after two cuts have been grouped into a subsample with a total exposure time of $\sim$1710 ks to the Crab pulsar, which are listed in Appendix \ref{app:tables}. }
	\label{fig:sign_vs_hp}
\end{figure}

We accumulate the data into each HP bins. The pulse profiles of two-hit events as a function of HP bins are shown in Figure \ref{fig:obs_vs_hp}, where one can observe how the statistics over HP bins differ from each other. In order to select a sub-sample of HP, we calculated the pulse significance of the Crab two-hit events for 400 HP bins by using equation (17) of \cite{1983ApJ...272..317L}. Figure \ref{fig:sign_vs_hp} shows the pulse significance as a function of HP. For the polarization analysis, we set a threshold value of 2 $\sigma$ on pulse significance (red dashed line in Figure \ref{fig:sign_vs_hp}). Additionally, we cut data with incident angle $\theta > 70^{\circ}$ (i.e., HP > 239), since the large off-axis incidence has worse polarization response as explained in Appendix \ref{app:systematic}. Finally, a sub-sample of 108 HP datasets for the Crab pulsar is used. These are listed in Table~\ref{table:observationODs_papd}, and the total exposure time of the subsample is $\sim$1710 ks. 
	
\begin{figure}
	\centering
	\includegraphics[width=7.3cm,height=7cm]{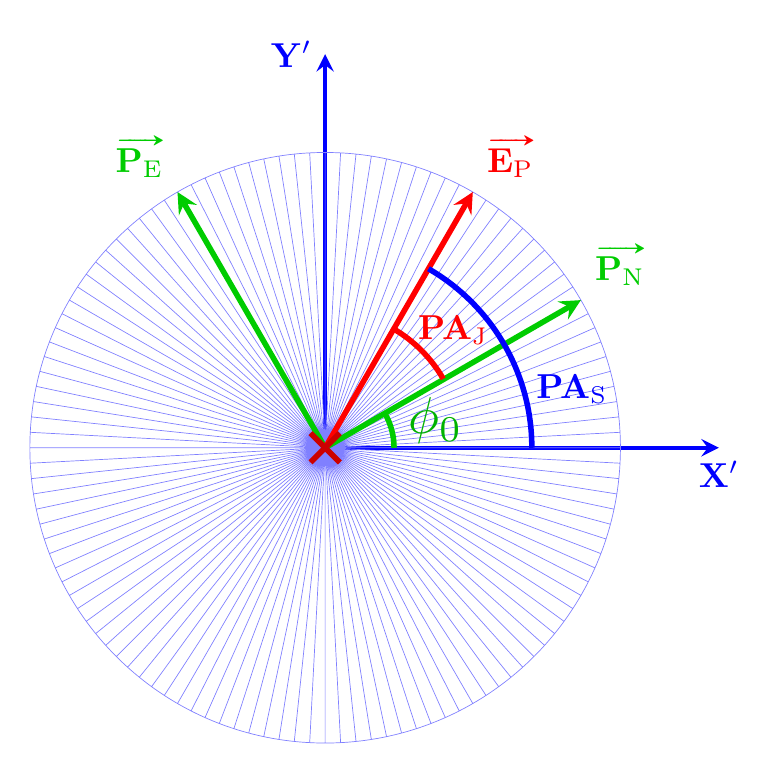}
	\caption{\textbf{Two PA definitions and the rotation of the source.} The view angle is from the source to POLAR. The X$'$-Y$'$ plane is that of the source coordinates. The \textbf{P$_{\textbf{N}}$}-\textbf{P$_{\textbf{E}}$} and \textbf{E$_{\textbf{P}}$} are also shown here. The local definition of PA in the source X$'$-Y$'$ plane is $\rm PA_{S}$, angle between \textbf{E$_{\textbf{P}}$} and \textbf{X}$'$. While the formal PA definition in J2000 North-East plane is $\rm PA_{J}$, the angle between \textbf{E$_{\textbf{P}}$} and \textbf{P$_{\textbf{N}}$} eastwards from the north. The difference between the two value is defined as the rotation angle $\phi_0$. The $\phi_0$ space is binned from 0-180$^{\circ}$ with a step size of 3$^{\circ}$.}
	\label{fig:projection_plane}
\end{figure}

Furthermore, for the Crab pulsar observation within one specific HP bin, the North-East plane of the J2000 coordinates onto the X$'$-Y$'$ plane of the source coordinates also change. In other words, the source rotates along the incident direction even when the local incident angles are fixed. Figure \ref{fig:projection_plane} shows the X$'$-Y$'$ plane of the source coordinates that are defined in Figure \ref{fig:coordinates}. The \textbf{E$_{\textbf{P}}$} is the electric vector of the source. The \textbf{P$_{\textbf{N}}$} and the \textbf{P$_{\textbf{E}}$} are the projection of the North and East vector of the J2000 coordinates respectively. The local definition of PA in source X$'$-Y$'$ plane is $\rm PA_{S}$, angle between \textbf{E$_{\textbf{P}}$} and \textbf{X}$'$. While the formal PA definition in the J2000 North-East plane is $\rm PA_{J}$, angle between \textbf{E$_{\textbf{P}}$} and \textbf{P$_{\textbf{N}}$} eastwards from the north. Hereby, we define the difference between the two value as rotation angle $\phi_0$.

\begin{figure}
	\centering
	\includegraphics[width=8cm,height=6cm]{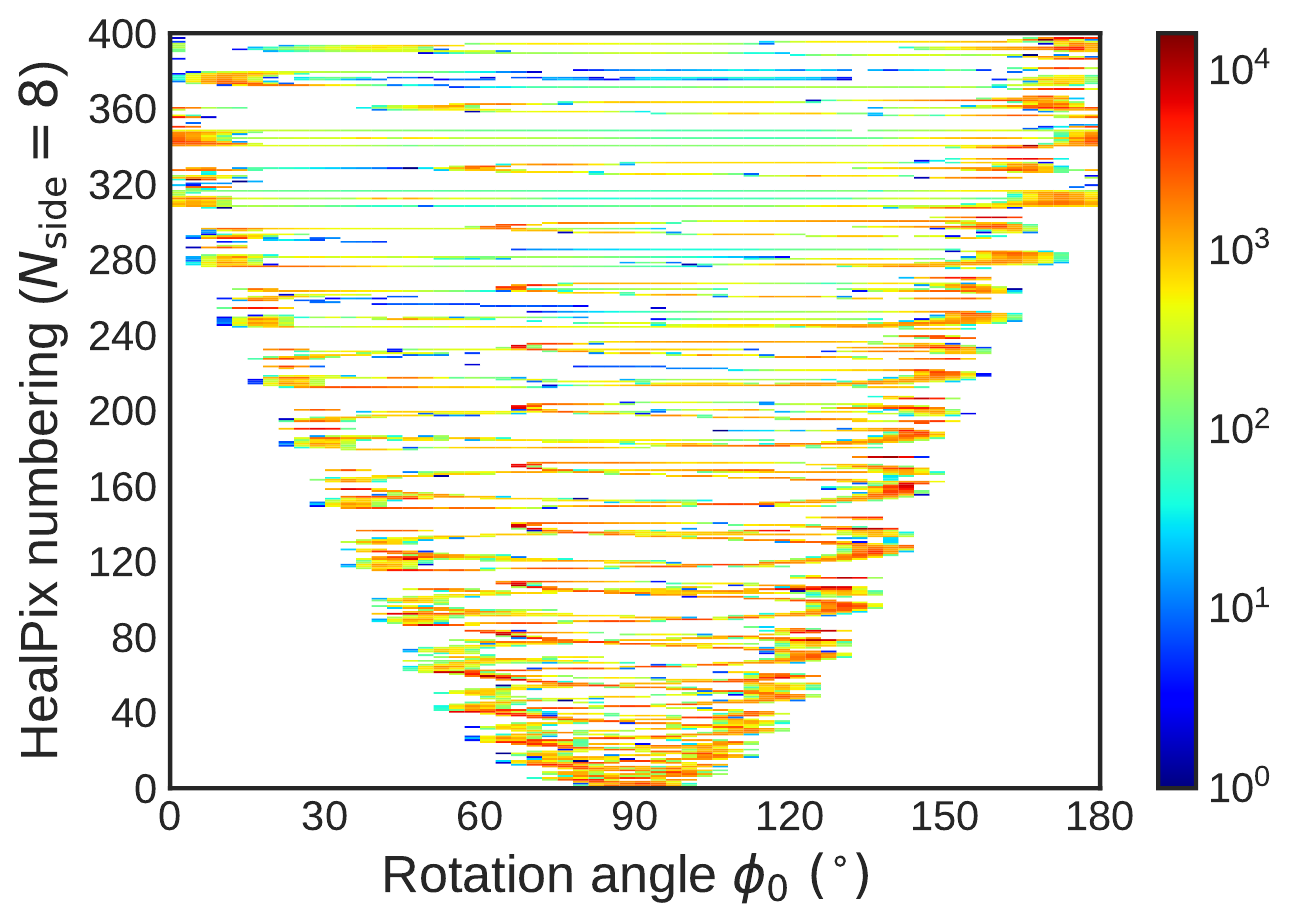}
	\caption{\textbf{The exposure time of the Crab pulsar v.s. healpix bin and $\phi_0$.} The Z-axis contains the exposure time of the Crab pulsar in unit of second. The exposure time has a parabolic distribution.}
	\label{fig:exposure_phi0}
\end{figure}

Based on the POLAR zenith pointing data, the value of $\phi_0$ for each arriving photon is calculated. We accumulate the data and bin $\phi_0$ from 0-180$^{\circ}$ with a step size of 3$^{\circ}$. The resulted exposure time of the Crab pulsar as a function of HP and $\phi_0$ are shown in Figure \ref{fig:exposure_phi0}, where one could clearly see a parabolic exposure time distribution. This variation of $\phi_0$ suggests that we cannot simply accumulate the two-hit into modulation curves for each HP, since the $\phi_0$ variation will counteract the modulation. We introduce a de-rotation method to align all $\phi_0$ bins to zero. In this case, the modulation curves of all $\phi_0$ bins can be stacked, and there will be no difference between $\rm PA_{J}$ and $\rm PA_{S}$. In Appendix~\ref{app:rotation} we show the de-rotation method, and check the validity of this method for polarization reconstruction. In Appendix~\ref{app:systematic}, we examine the systematic effect of the polarization reconstruction versus rotation angle $\phi_0$ and incident angle $\theta$. Thereafter, based on the remaining HP subsample and the de-rotation method, we generate 108$\times$3 modulation curves for 108 HP and 3 phase intervals.
	
%-------------------------------------------------------------------------------------------
\section{Methodology}\label{sec:methodology}

Below in Section \ref{subsec:con_rec}, we will describe the conventional polarization analysis method as a general introduction to gamma-ray polarimetry. We only use this method in Appendix \ref{app:rotation} and Appendix \ref{app:systematic} for a quick check on the PA and PD values. However, we did not use this method for the final polarization analysis in \cite{2019NatAs...3..258Z} and neither in this paper; the methods used in the latter two works are presented in Section \ref{subsec:polar_grb} and \ref{subsec:polar_crab}.

\subsection{Conventional polarization reconstruction}\label{subsec:con_rec}

\begin{figure}
    \centering
    \includegraphics[width=6.5cm,height=7cm]{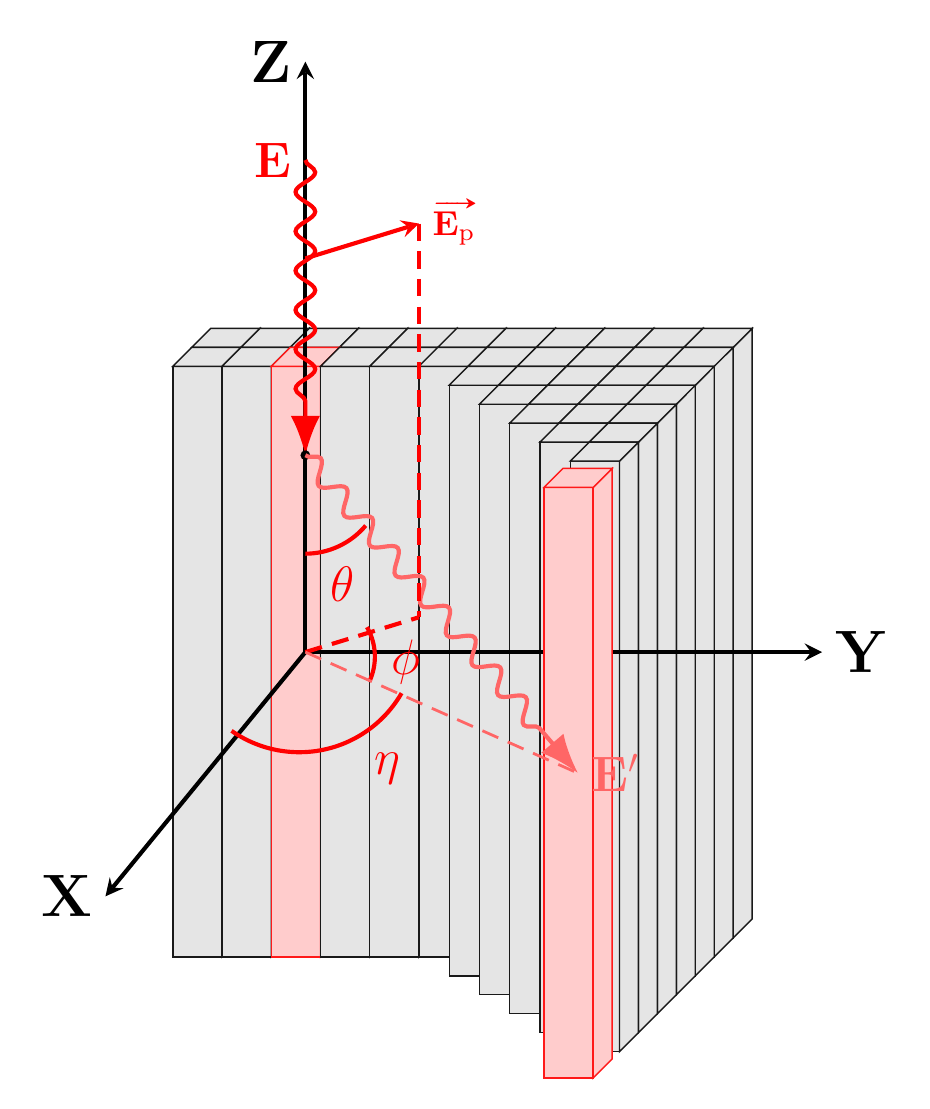}
    \caption{\textbf{Illustration of the Compton scattering process in a polarimeter.} The incident photon with an energy $E$ Compton-scatters in one detector segment before moving towards a second detector segment with energy $E'$, where it interacts a second time. The polar scattering angle $\theta$ is shown together with the azimuthal scattering angle $\eta$, both of which are measured in the detector. Angle $\phi$, which is defined as the angle between the intrinsic polarization vector and the momentum vector in the X-Y plane after Compton scattering, is preferentially $90^\circ$ and suppressed at $0^\circ$.}
    \label{fig:scattering} 
\end{figure}

The Compton scattering process of photons and free electrons follows the Klein-Nishina differential cross-section equation \citep{1929ZPhy...52..853K}:

\begin{equation}\label{equ:KN_equ}
\frac{d\sigma_C}{d\Omega} = \frac{r_{\rm e}^2}{2}\ \varepsilon^2 \left\{ \varepsilon+\varepsilon^{-1}-\sin^2\theta+\sin^2\theta \cos\left[2\left(\eta+\frac{\pi}{2}\right)\right] \right\} \ ,
\end{equation}

where $r_{\rm e}$ is the classical electron radius; $\varepsilon$ is the energy ratio of the outgoing photon over the energy of the incident photon, that is, $E'/E$; $\theta$ is the angle between the photon's incident and outgoing direction; $\eta$ is defined as the azimuthal angle on the X-Y plane perpendicular to photon incident directions. Here we take the incidence direction to be along the detector's zenith as an example to discuss the polarization reconstruction. For a small off-axis angle, the situation is similar, however for a large off-axis angle there are additional effects which are discussed in section Appendix~\ref{app:systematic}. The scattering process is illustrated in Figure \ref{fig:scattering}.
    
By integrating equation (\ref{equ:KN_equ}) over the solid angle, we have:

\begin{equation}\label{equ:sigma_integration}
\sigma_C = 2\pi \int_{0}^{\pi} \frac{d \sigma_C}{d \Omega} \sin \theta d\theta \ .
\end{equation}

For a photon with an initial energy $E$ and PA of $\Phi$ will, after scattering with free electrons, have a probability distribution for the azimuthal angle of the outgoing $\eta$ :
\begin{equation}\label{equ:prob_modu}
P(\eta,\ E,\ \Phi) = \left\{a(E)\times\cos\left[2\left(\eta - \Phi + \frac{\pi}{2}\right)\right] + b(E)\right\}\times c(E) \ ,
\end{equation}

where $a(E)$, $b(E)$ and $c(E)$ are constants related to energy, and $c(E)$ is the probability normalization factor. Assuming that a beam of photons is polarized, with a specific PD and PA, $\eta$ of the outgoing photons should follow the trigonometrical probability function with the same phase. So the final scattering angle distribution (often called the modulation curve) accumulated over events, that is, the histogram of $\eta$, should have the equivalent form of equation (\ref{equ:prob_modu}), written as:

\begin{equation}\label{equ:counts_modu}
C(\eta,\ \Phi) = A\times\cos\left[2\left(\eta - \Phi + \frac{\pi}{2}\right)\right] + B \ .
\end{equation}
    
In case the beam is unpolarized, photons have different PA values, which result in different phases of $\eta$'s trigonometrical probability function that will cancel out the total modulation, so we expect the modulation curve to be flat.
    
By fitting the modulation curve with equation (\ref{equ:counts_modu}), we can obtain the relevant parameter values, where PA (=$\Phi$) can be directly obtained by fitting. As for PD, we define the modulation factor:

\begin{equation}\label{equ:mu_factor}
\mu = \frac{A}{B} = \frac{C_{\rm max}-C_{\rm min}}{C_{\rm max}+C_{\rm min}} \ .
\end{equation}
    
The modulation factor $\mu$ can be obtained by fitting the measured modulation curve with equation (\ref{equ:mu_factor}). The modulation factor $\mu$ reflects the PD of linear polarization of the photons. Generally, a polarimeter can be calibrated with a 100\% linearly polarized source and verified by Monte Carlo simulation to obtain $\mu_{100}$. Then the PD of a measurement can be calculated by:

\begin{equation}\label{equ:pd_value}
{\rm PD} = \frac{\mu}{\mu_{100}} \ .
\end{equation}
    
The above steps are for ideal conditions. In practice, a detector has geometric effects that may contaminate the modulation curve. It is therefore necessary to correct for these effects before fitting the modulation curve. While integrating equation (\ref{equ:KN_equ}), the geometric effect $G(\eta,\ \theta_{\gamma})$ needs to be taken into account, where $\theta_{\gamma}$ is the incident angle of the source with respect to the detector. So for a polarized (PD = $\Pi$) beam:

\begin{equation}\label{equ:pol_integral}
C_{\Pi}\ (\eta,\ \Phi,\ \theta_{\gamma}) = G(\eta,\ \theta_{\gamma})\cdot 2\pi\int_{0}^{\pi} \frac{d \sigma_{C}^{\ \Pi}}{d \Omega} \sin \theta d\theta \ .
\end{equation}
    
Correspondingly, for a unpolarized (PD = 0) beam (usually obtained by Monte Carlo simulation):

\begin{equation}\label{equ:unp_integral}
C_0\ (\eta,\ \Phi,\ \theta_{\gamma}) = G(\eta,\ \theta_{\gamma})\cdot 2\pi\int_{0}^{\pi} \frac{d \sigma_{C}^{\ 0}}{d \Omega} \sin \theta d\theta \ .
\end{equation}
    
Therefore, the geometric effect can be eliminated by the following correction to obtain a geometric-corrected modulation curve:

\begin{equation}\label{equ:pol_unp}
C(\eta,\ \Phi,\ \theta_{\gamma})=\frac{C_{\Pi}\ (\eta,\ \Phi,\ \theta_{\gamma})}{C_0\ (\eta,\ \Phi,\ \theta_{\gamma})} \cdot C_{\rm norm} \ ,
\end{equation}

where $C_{\rm norm}$ is a normalization factor. Both measured and simulated modulation curves can be modified in this way for geometric effect correction. After that, by fitting $C(\eta,\ \Phi,\ \theta_{\gamma})$ with equation (\ref{equ:counts_modu}), and then calculate according to the equation (\ref{equ:mu_factor}) and (\ref{equ:pd_value}), PD and PA can be reconstructed. The parameter errors can be acquired by fitting the measured and simulated modulation curves and carefully handling their error propagation. 

\subsection{POLAR GRB polarization reconstruction}\label{subsec:polar_grb}
    
\begin{figure}
    \centering
    \includegraphics[width=8.5cm,height=6cm]{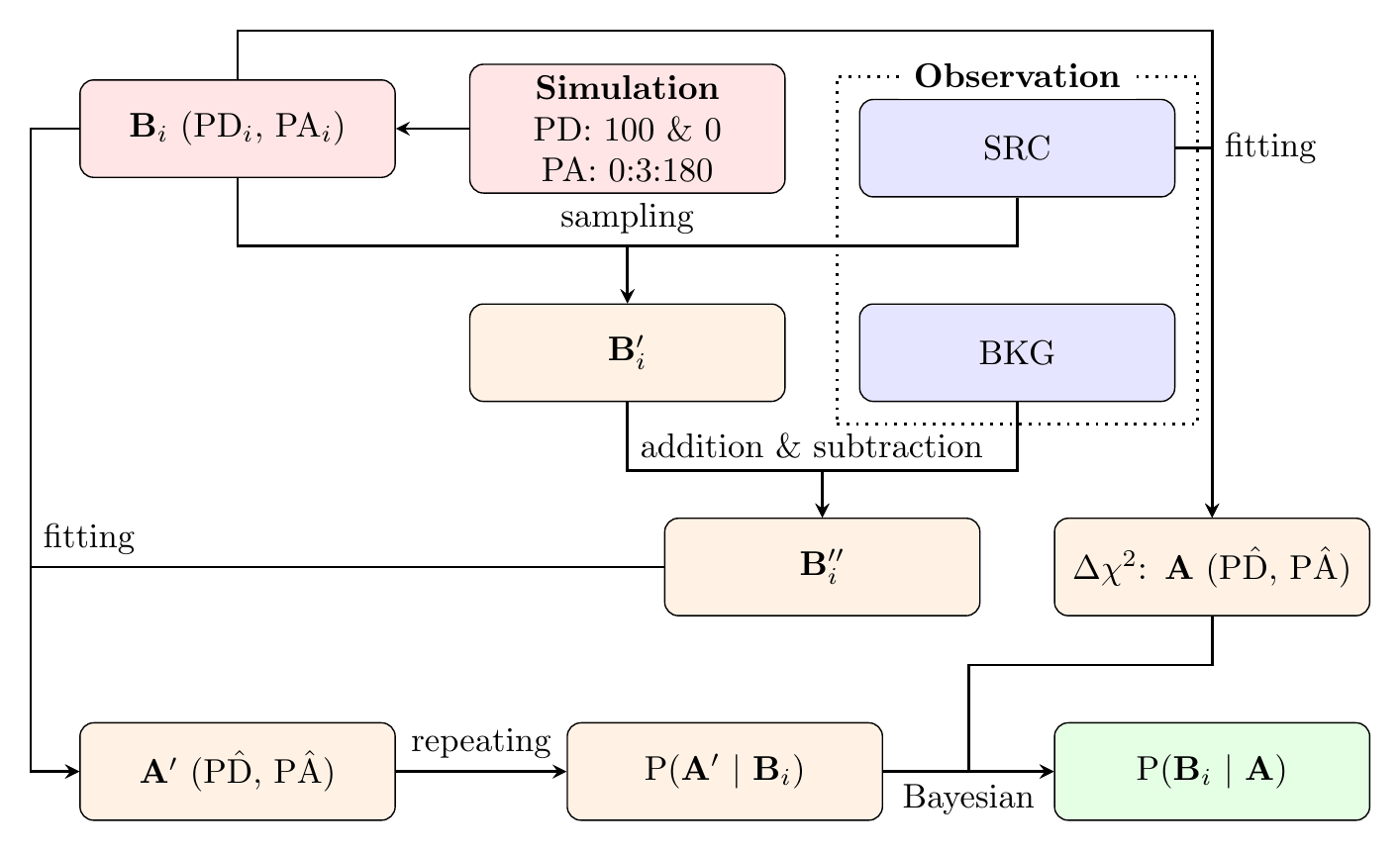}
    \caption{\textbf{Polarization Bayesian analysis workflow for GRB analysis.} This workflow is summarized from \protect\cite{2019NatAs...3..258Z}. Finally the posterior written as P(\textbf{B$_i$} | \textbf{A}) on the bottom right is generated for polarization reconstruction.}
    \label{fig:pdpa_method_grb}
\end{figure}

The first polarization results of a subsample of 5 GRBs detected by POLAR were published in \cite{2019NatAs...3..258Z}. The analysis workflow presented there is different from (but equivalent to) the one described in Section \ref{subsec:con_rec}. The differences are: 1) the geometric effect of a measured modulation curve will not be corrected with equation (\ref{equ:pol_unp}) since the Monte Carlo simulations include the same geometric effects; 2) the PA \& PD are directly reconstructed by fitting measured modulation curves against a range of simulated modulation curves with different PD and PA. Subsequently, PD and PA are found using the least $\chi^2$ method. Furthermore, \cite{2019NatAs...3..258Z} has introduced a Bayesian approach for PA and PD reconstruction. The whole workflow is shown in Figure \ref{fig:pdpa_method_grb}. Hereby we briefly introduce the analysis workflow:
    
\begin{itemize}

    \item \textbf{Observation}: the GRB source interval and background intervals before and after the GRB  are carefully defined. Then two-hit events are selected based on \cite{2018NIMPA.900....8L} and the scattering angles for these events are calculated to build modulation curves. The BKG (background) modulation curve is scaled with equivalent live time of the GRB interval, and the BKG modulation curve is subtracted from the SRC (source) modulation curve;
    	
    \item \textbf{Simulation}: the MC software \citep{2017NIMPA.872...28K} is initially used to perform a set of 61 simulations (PD 100\% and PA 0:3:180$^{\circ}$ (0 to 180$^{\circ}$ with step size of 3$^{\circ}$); PD 0\% and PA 0$^{\circ}$) for the GRB using an energy spectrum and localization provided by other instruments. The 61 simulated data are processed using the same analysis pipeline used for the measurement data which could be referred to \citep{2018NIMPA.900....8L};
    	
    \item \textbf{B$_i$}: simulated modulation curves for different PD$_i$ and PA$_i$ (101$\times$61 sets in total, PD: 0:1:100\%, PA: 0:3:180$^{\circ}$) are derived from the initial simulation set described above, by mixing those of the unpolarized flux and the 100\% polarized flux (PA$_i$) with the corresponding weights of (100\%-PD$_i$) and PD$_i$ respectively;
    	
    \item $\Delta\chi^2$: by fitting directly the measured (BKG subtracted) modulation curve with all the 6161 simulated ones the $\chi^2$ values of the 6161 fits is retrieved. These are used to produce a $\Delta\chi^2$ distribution in PD and PA space. Finally, the best fitted value \textbf{A} (P\^D, P\^A) together with 1-2-3 sigma confidence intervals are deduced from $\Delta\chi^2$ distribution;
    	
    \item \textbf{B$_i'$}: we sample the modulation curves with a known PD and PA from \textbf{B$_i$}, using the statistical quantity of the measured SRC modulation curves; 
    	
    \item \textbf{B$_i''$}: we introduce the BKG statistical fluctuations to \textbf{B$_i'$} by adding and subtracting of sampled BKG modulation curves (each with the number of counts from the data), thus we obtain 'the measurement simulations' with known polarization parameters;
    	
    \item P(\textbf{A$'$} | \textbf{B$_i$}): by fitting the \textbf{B$_i''$} equal as described before in the $\Delta\chi^2$ process, a best fitted value \textbf{A$'$} (P\^D, P\^A) for a given \textbf{B$_i$} will be recorded; this process is repeated for a statistically reasonable times (eg. 10000) to obtain a distribution of A. After normalizing this distribution to 1, we have the conditional probability P(\textbf{A$'$} | \textbf{B$_i$}), which indicates how likely A will be measured with a given B;
    	
    \item P(\textbf{B$_i$} | \textbf{A}): with respect to the best fitted value \textbf{A} (P\^D, P\^A) from $\Delta\chi^2$ approach, we calculate the posterior distribution by assuming a prior distribution which is uniformly distributed within physical polarization space (PD 0-100\% \& PA 0-180$^{\circ}$) and vanishes out of this space. 
    	
\end{itemize}
    
\begin{figure}
	\centering
	\includegraphics[width=8.5cm,height=6.5cm]{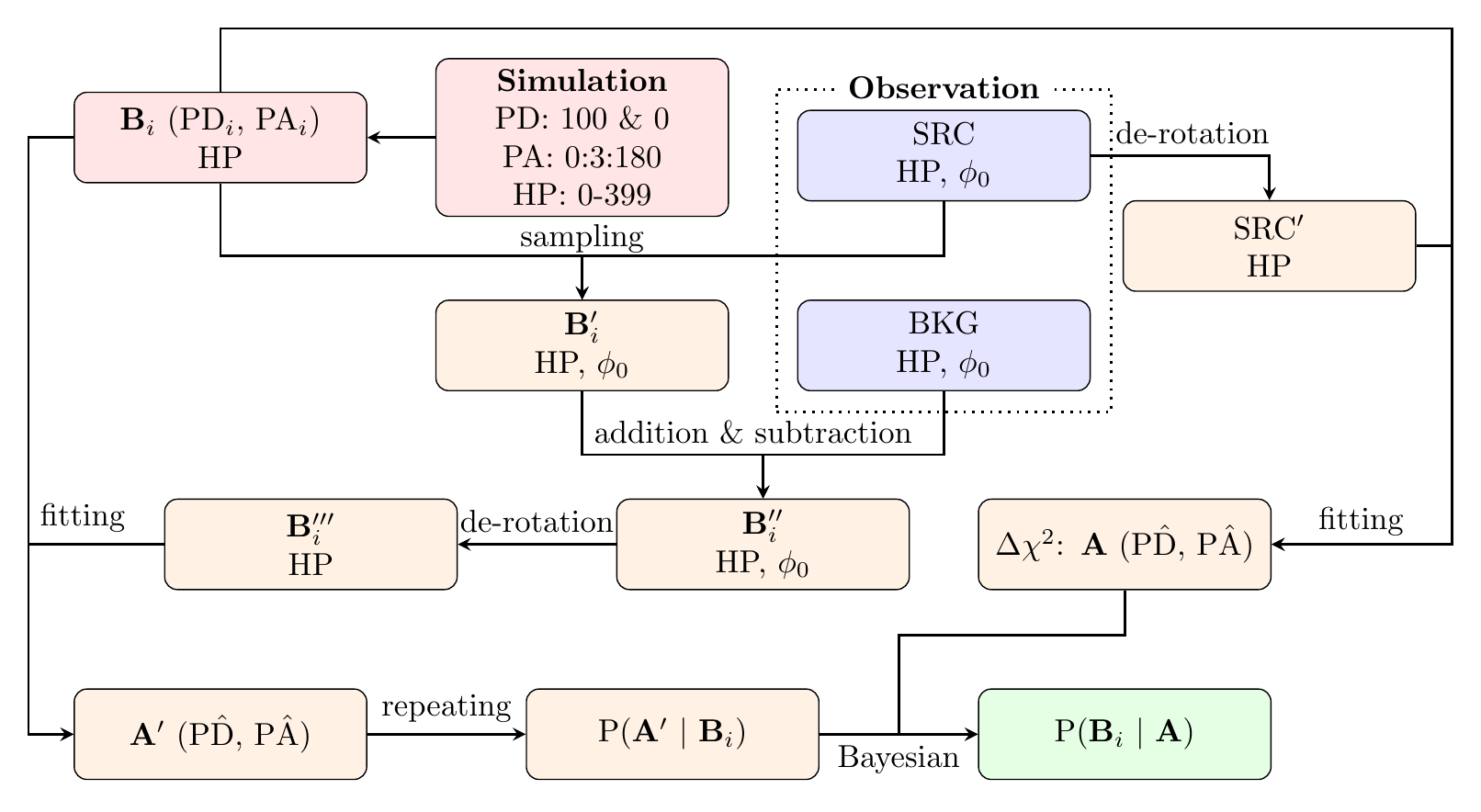}
	\caption{\textbf{Polarization Bayesian analysis workflow for the Crab analysis.} We adapted the GRB workflow to the Crab workflow with taking into account two more observational parameters which are the incident angle (reconstructed to healpix (HP) bin) and the rotation angle ($\phi_0$). The $\phi_0$ is corrected for using the de-rotation process, and then observation datasets in different HP bins are joint-fitted with simulation ones.}
	\label{fig:pdpa_method_crab}
\end{figure}
    
\subsection{POLAR Crab polarization reconstruction}\label{subsec:polar_crab}
	
The Crab observations have two more observational parameters compared to the GRB analysis. These are the incident angle (HP) and the rotation angle ($\phi_0$). So we adapt the GRB workflow to the Crab workflow as follows (only differences are shown, other parts are similar to the GRB workflow) which is also shown in Figure \ref{fig:pdpa_method_crab}:
	
\begin{itemize}

	\item \textbf{Observation}: for each (HP, $\phi_0$) dataset, there are three intervals defined as AP (0.0-1.0), P1 (0.0-0.2 || 0.8-1.0) and P2 (0.2-0.6). The background interval is defined as the 0.6-0.8 phase range. The BKG modulation curves are scaled with equivalent phase length to the different on-pulse intervals, and the BKG modulation curve is subtracted from the SRC modulation curve. Note that the BKG contains the nebula and background, so there will be no nebula contribution in the remaining SRC curve;
		
	\item \textbf{Simulation}: a set of 61$\times$400 (PD 100\% and PA 0:3:180$^{\circ}$; PD 0\% and PA 0$^{\circ}$; HP: 0-399) will be performed by using the spectrum of all phase intervals from \cite{2001A&A...378..918K}, we have added a 4\% relative error to each bin in the simulated modulation curves in order to take in to account the systematic uncertainties likely caused by spectral impact, similar to how this was done in \cite{2019NatAs...3..258Z}. As for $\phi_0$'s parameter space, it can be derived by shifting the curve uing the $\phi_0$ values in PA space;
		
	\item De-rotation: before fitting the measured modulation curves with simulated ones, we have to firstly accomplish a de-rotation process in order to fix $\phi_0$ which can be found in detail from Appendix~\ref{app:rotation}. Note that, during this process, both the SRC and BKG modulation curves need to be corrected for geometric effect using the unpolarized simulated modulation curve of this dataset in order to allow for the de-rotation process to work;
		
    \item SRC$'$: the de-rotation process is applied from SRC to SRC$'$, then $\phi_0$ parameter is fixed, and for each HP there will be only one modulation curve;
		
	\item \textbf{B$_i'''$}: from \textbf{B$_i''$} to \textbf{B$_i'''$} is same as from SRC to SRC$'$;
		
	\item $\Delta\chi^2$: the fitting will be jointly among different HP bins, and a $\Delta\chi^2$ distribution is obtained.
		
\end{itemize}
	
%-------------------------------------------------------------------------------------------
\section{Results and discussion}\label{sec:result}
	
\begin{figure}
	\centering
	\includegraphics[width=8cm,height=7cm]{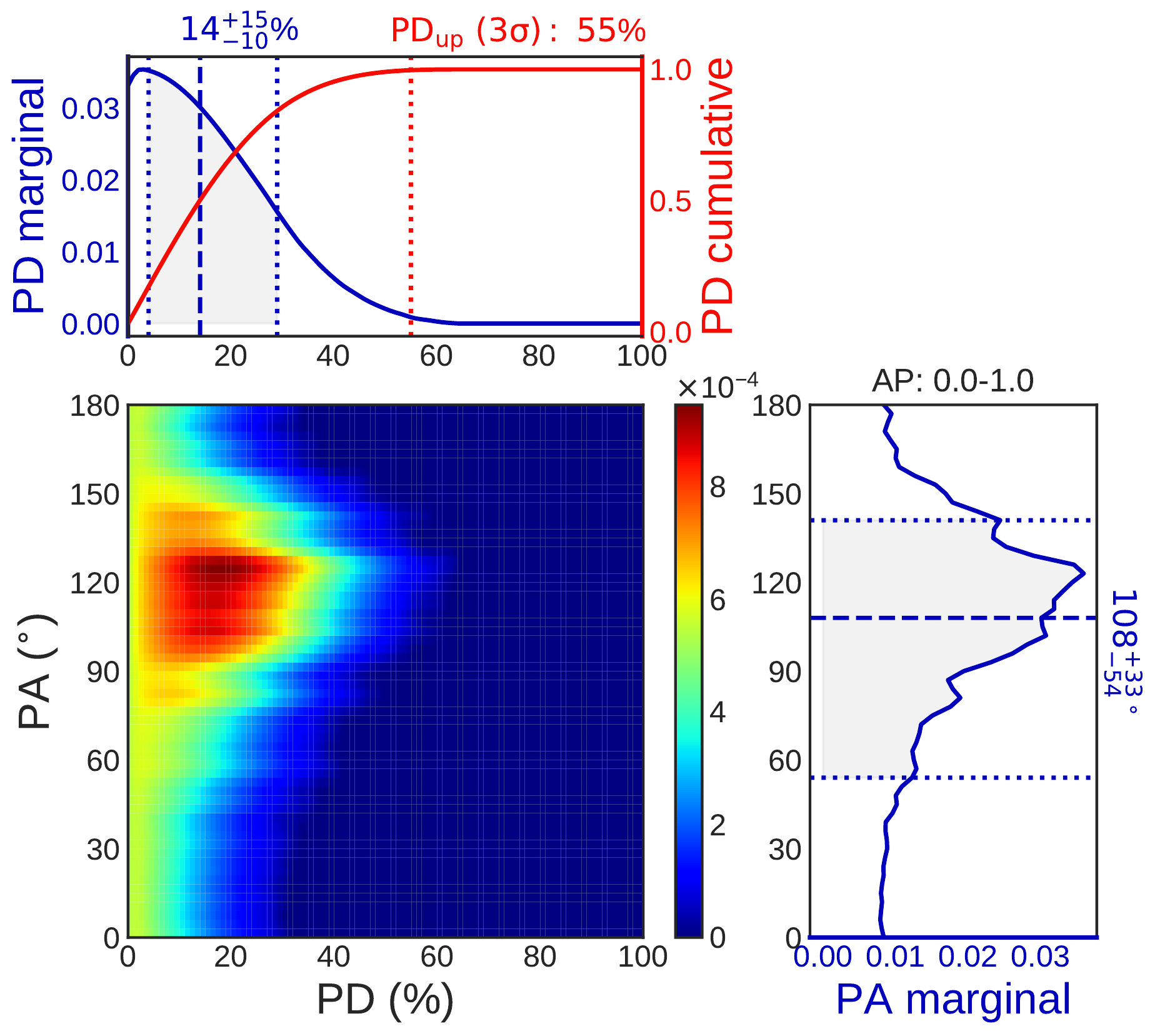}
	\caption{\textbf{PD/PA posterior of AP phase interval.} The best fitted values and the corresponding 1$\sigma$ deviations are PD=$14^{+15}_{-10}$\%, PA=$108^{+33}_{-54}$$^{\circ}$, the 3$\sigma$ upper limits on PD is 55\%, and the probability that true PD is smaller than 1\% is 3.31\%.}
	\label{fig:pdpa_post_ap} 
\end{figure}

\begin{figure}
	\centering
	\includegraphics[width=8cm,height=7cm]{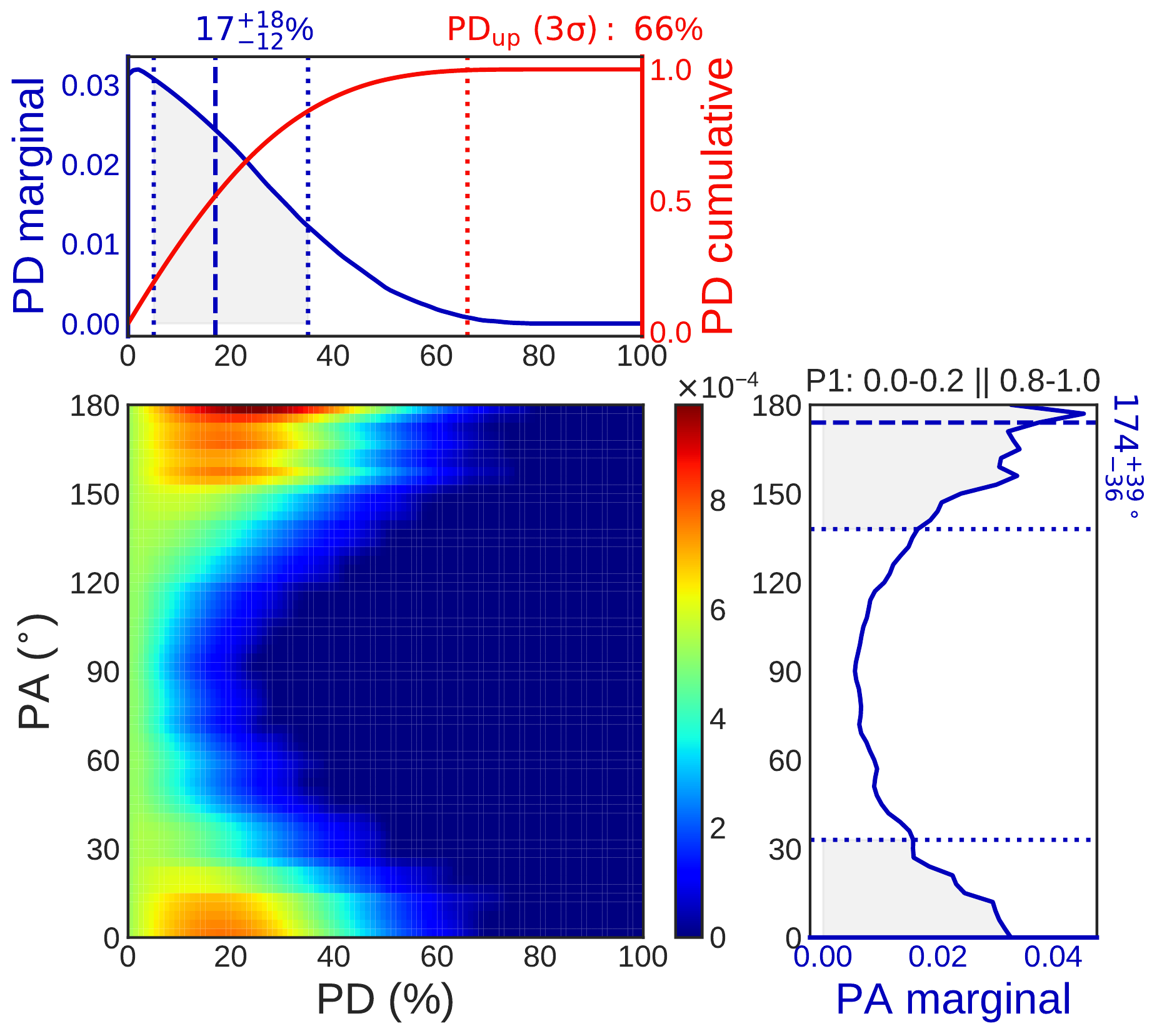}	
	\caption{\textbf{PD/PA posterior of P1 phase interval.} The best fitted values and the corresponding 1$\sigma$ deviations are PD=$17^{+18}_{-12}$\%, PA=$174^{+39}_{-36}$$^{\circ}$, the 3$\sigma$ upper limits on PD is 66\%, and the probability that true PD is smaller than 1\% is 3.13\%.}
	\label{fig:pdpa_post_p1} 
\end{figure}

\begin{figure}
	\centering
	\includegraphics[width=8cm,height=7cm]{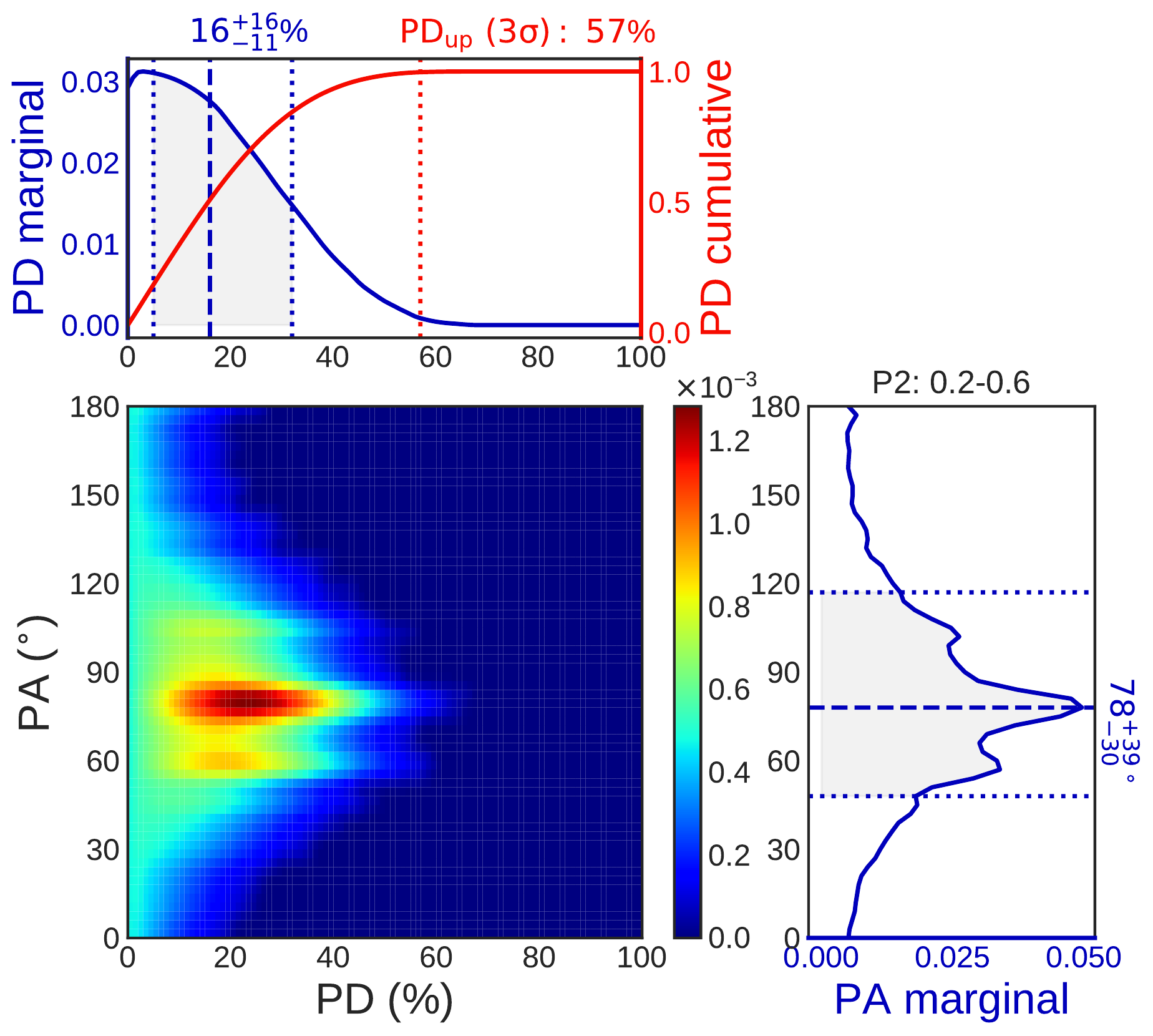}	
	\caption{\textbf{PD/PA posterior of P2 phase interval.} The best fitted values and the corresponding 1$\sigma$ deviations are PD=$16^{+16}_{-11}$\%, PA=$78^{+39}_{-30}$$^{\circ}$, the 3$\sigma$ upper limits on PD is 57\%, and the probability that true PD is smaller than 1\% is 2.92\%.}
	\label{fig:pdpa_post_p2} 
\end{figure}

Based on the methods described in the previous section, we obtain polarization measurements for the Crab pulsar. The posterior distributions are shown in Figure \ref{fig:pdpa_post_ap}, Figure \ref{fig:pdpa_post_p1} and Figure \ref{fig:pdpa_post_p2} respectively. The best fitted values and the corresponding 1$\sigma$ deviations are AP (PD=$14^{+15}_{-10}$\%, PA=$108\substack{+33 \\ -54}^{\circ}$), P1 (PD=$17\substack{+18 \\ -12}$\%, PA=$174\substack{+39 \\ -36}^{\circ}$) and P2 (PD=$16\substack{+16 \\ -11}$\%, PA=$78\substack{+39 \\ -30}^{\circ}$). From the cumulative probability of PD, we obtain the 3$\sigma$ upper limits on PD are AP (55\%), P1 (66\%) and P2 (57\%) respectively, where the upper limits of P1 and P2 are more stringent than previous results and all of them reject a high PD scenario. Furthermore, the probability that the true PD is smaller than 1\% for AP, P1 and P2 are 3.31\%, 3.13\% and 2.92\% respectively, which, although not being a significant exclusion, suggest that the Crab pulsar emission is unlikely unpolarized. All the results of this work, together with previous work, are listed in Table~\ref{table:PDPA_resultstable}. 
	
\begin{figure}
	\centering
    \includegraphics[width=7cm,height=6cm]{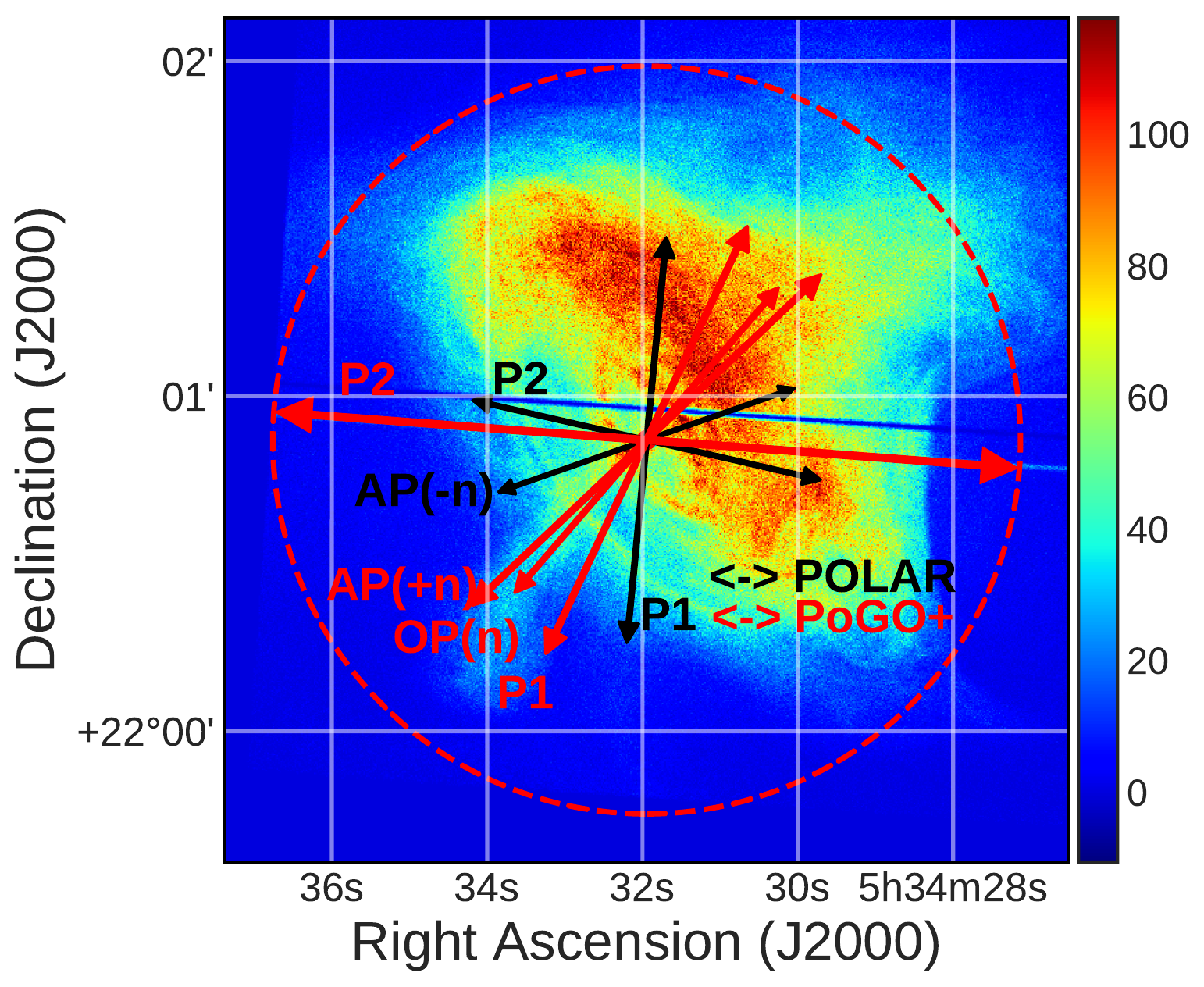}
    \caption{\textbf{Polarization vectors projected onto the Crab image taken by Chandra.} Directions of the vectors represent the PA values counting from the top (North) to the left (East). The lengths of the vectors are scaled according to the PD such that the longest one corresponds to 33.5\% (the dashed circle, PD of P2 by PoGO+). The polarization vectors for AP, P1 and P2 as obtained from the POLAR analysis are shown in red while those of PoGO+ are, for comparison, shown in black. The AP from PoGO+ contains the nebula contribution (represented by the OP interval). The estimated 20.0$\pm$24.8\% by maximizing a posterior probability is used for the PD of P1 by PoGO+. Note that this figure does not visualize the errors on PD and PA, which can be found in Table \ref{table:PDPA_resultstable}.} 
    \label{fig:pa_onto_crab} 
\end{figure}

We take PoGO+ as an example for detailed comparison. First of all, our results are in good agreement with each other within the margins of errors. Then, in order to have a direct impression, we plot the polarization vectors onto the Crab X-ray image taken with Chandra\footnote{\url{https://chandra.harvard.edu/photo/2009/crab/fits}} in Figure \ref{fig:pa_onto_crab}. The black vectors in this plot are of POLAR, the red ones are for PoGO+. The cross point of these vectors is the Crab pulsar. Directions of the vectors are represented for the PA values counting from the top (North) to the left (East), and the lengths of the vectors are scaled by 33.5\% (the dashed circle, PD of P2 by PoGO+) for PD values. For the polarization result of P1 by PoGO+, the marginalized posterior estimate (Marg.) is PD=$0^{+29}_{-0}$\% and PA=$153 \pm 43^{\circ}$, but the maximum a posterior probability estimate (MAP) gives PD=20.0$\pm$24.8\% and PA=$153 \pm 36^{\circ}$ \citep{2017NatSR...7.7816C}. As a zero PD value (Marg. result) is unmeasurable (positive-definite) and makes also PA value meaningless, the MAP result (PD=20.0$\pm$-24.8\%, PA=$153 \pm 36^{\circ}$) is used in Figure \ref{fig:pa_onto_crab} to demonstrate the polarization vector. The POLAR results of AP, P1 and P2 do not include the nebula contribution. The OP result of PoGO+ is contributed purely by the nebula, the PA of which is parallel to the spin axis of the Crab pulsar. The AP polarization results from PoGO+ contain also a nebula contribution which is more dominant than that of the pulsar, such that the fitted PA value is significantly different from that of POLAR. Due to the limited statistics, the phase intervals of P1 and P2 in POLAR are generally wider than those in PoGO+, which means that POLAR results include more components than that of PoGO+. The optical phase-resolved polarization results suggest that the PDs are higher in other phase ranges (especially in bridge and off-pulse) than those in the two narrow peaks, and the PAs swing through a large angle in both peaks. If the same polarization properties are similar at high energies, polarization result of a wider phase range should be different from that of a narrower one. This might explain the differences of PAs in P1 and P2 between POLAR and PoGO+. We should note here that as Figure \ref{fig:pa_onto_crab} does not visualize the errors on PD and PA (can be found in Table \ref{table:PDPA_resultstable}), the attempt to interpret the differences between POLAR and PoGO+ applies only to the best fitted values without taking into account the margins of errors. In any case, the results on different definitions of phase intervals are difficult to compare, and we expect a finer phase-resolved polarization measurements with the same phase-intervals as those in optical bands will be achieved by future missions.
	
%-------------------------------------------------------------------------------------------
\section{POLAR-2}\label{sec:polar-2}

POLAR-2 is a follow-up of the POLAR mission, which has been already selected to be launched to the China Space Station in around 2024 and will operate at least for two years \citep{2019ICRC...36..572K, 2021arXiv210902977K,2020SPIE11444E..2VH, 2021arXiv210902978D}. In order to improve the polarimetry precision, the POLAR-2 design takes two major improvements compared to POLAR, which are: 1) increasing the sensitive detector (scintillator array) modules from 25 to 100 (each module has 64 independent scintillator bars) so that the effective area will be approximately four times larger; 2) using a SiPM array (64 channels per module) together with its own front-end electronics (FEE) for readout which will lower the energy threshold of the detector pixel and increase the effective area at low energies. These features increase the SNR of a source detection significantly. Based on the preliminary design of POLAR-2, Monte-Carlo simulations have been performed to obtain the effective area as a function of energy. Based on \cite{2019ICRC...36..572K} and \cite{2019JHEAp..24...15L}, we consider the amount of measured photons with POLAR-2 is approximately 10 times that of POLAR in the 50-500 keV energy range by the convolution of the effective area and the Crab pulsar spectrum. Since the amount of background photons is expected to be directly proportional to the geometrical area of the instrument, we can conservatively assume the POLAR-2 background level to be 4 times that of POLAR. Detailed background simulations are however required in the future to provide a more accurate number for this.
    
\begin{figure}
    \centering
    \includegraphics[width=8cm,height=14cm]{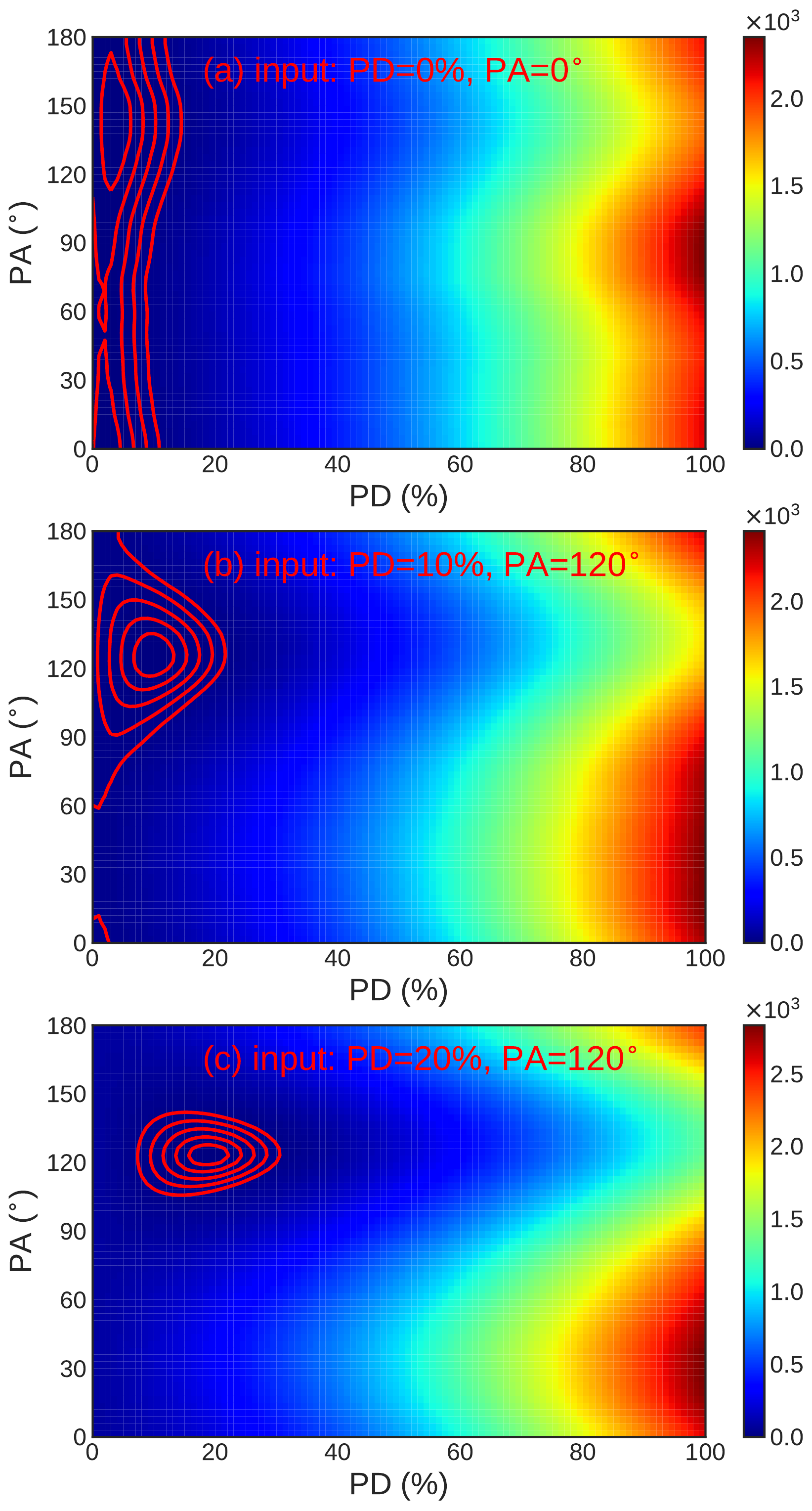}	
    \caption{\textbf{Simulated POLAR-2 polarization reconstruction for the AP interval of the Crab pulsar with 2 years observation.} The $\Delta\chi^2$ contour maps from top to bottom are for given polarization values of (PD=0\%, PA=0$^{\circ}$), (PD=10\%, PA=120$^{\circ}$) and (PD=20\%, PA=120$^{\circ}$) respectively. The contour lines from inner to outer are $\chi^2$ (degree of freedom is 2) values of 2.3, 6.18, 11.83, 19.33 and 28.74, representing 1-5 sigma confidence levels respectively.}
    \label{fig:polar2-ap} 
\end{figure}

Based on POLAR's polarization reconstruction workflow for Crab introduced in Section \ref{subsec:polar_crab}, we perform the simulated polarization reconstruction of 2 years observation of the Crab pulsar for given polarization values, which is then joint-fitted among \textbf{B}$_i$ (PD$_i$, PA$_i$, HP) and \textbf{B}$'''_i$ (HP). For the AP phase interval of the Crab pulsar, the $\Delta\chi^2$ contour maps of given polarization values of (PD=0\%, PA=0$^{\circ}$), (PD=10\%, PA=120$^{\circ}$) and (PD=20\%, PA=120$^{\circ}$) are shown in Figure \ref{fig:polar2-ap}(a), \ref{fig:polar2-ap}(b) and \ref{fig:polar2-ap}(c) respectively. The contour lines from inner to outer are $\chi^2$ (degree of freedom is 2) values of 2.3, 6.18, 11.83, 19.33 and 28.74, representing 1-5 sigma confidence levels respectively. Within 2 years of observation with POLAR-2, if the Crab pulsar is unpolarized, POLAR-2 will give 5 sigma upper limits on PD to a level of $\sim$20\%. While if the Crab pulsar is 10\% or 20\% polarized, POLAR-2 will be able to confirm the Crab pulsar to be polarized with $4\sigma$ or $5\sigma$ confidence level. 
    
%-------------------------------------------------------------------------------------------
\section{Summary}\label{sec:summary}

We present here the polarization measurement results of the AP, P1 and P2 phase intervals of the Crab pulsar performed using POLAR data. The measurements are in good agreement with previous results such as those from PoGO+. As POLAR can identify the pulsar, opposed to PoGO+ which can also investigate the nebula signal, it was capable of producing the most strict 3$\sigma$ PD upper limits of P1 (66\%) and P2 (57\%). All of these results reject a high PD scenario. Furthermore, the probability that the true PD is smaller than 1\% for AP, P1 and P2 are 3.31\%, 3.13\% and 2.92\% respectively, which, although not being a significant exclusion, show that the Crab pulsar emission is unlikely unpolarized. Although the polarization measurements of this work are lacking statistical precision, the polarimetry methodology developed here could be employed for other wide FoV polarimeters and thereby increases the scientific targets available to these instruments from transient sources to pulsars. Using the polarimetry methodology and Monte-Carlo simulations we extrapolate the polarization reconstruction ability of POLAR-2 with 2 years observation to the Crab pulsar. We can therefore conclude that with 2 years of observation to the Crab pulsar POLAR-2 will be able to confirm the emission to be polarized with $5\sigma$ confidence level in case the polarization is $20\%$, while it can be done with $4\sigma$ confidence level if it is polarized at $10\%$. In case the emission is unpolarized, POLAR-2 could obtain $5\sigma$ upper limits at a level of $\sim$20\%. With sufficient statistics of POLAR-2, a finer phase-resolved polarization measurements of the Crab pulsar can also be obtained. Additionally, as POLAR-2 will have an increased energy range, spanning from 30 keV to 800 keV, energy dependent polarization measurements will also be within the possibilities. Such measurements can be combined with those from photoelectron-track polarimeters onboard the IXPE mission \citep{2021AJ....162..208S} and the eXTP mission \citep{2019SCPMA..6229502Z} in the near future. At the time of the launch of POLAR-2, IXPE is expected to have performed precise measurements of the Crab emission in the 2-10 keV energy range, while eXTP will likely increase the precision during, or after, the POLAR-2 mission.

\section*{Acknowledgments}\label{sec:ackonwledgment}
	
We gratefully acknowledge the financial support from the National Natural Science Foundation of China (Grant No. 11961141013, 11503028), the Joint Research Fund in Astronomy under the cooperative agreement between the National Natural Science Foundation of China and the Chinese Academy of Sciences (Grant No. U1631242), the Strategic Priority Research Program of the Chinese Academy of Sciences (Grant No. XDB23040400), the National Natural Science Foundation of China (Grants U1838201, U1838202 and U1838104), the Youth Innovation Promotion Association of Chinese Academy of Sciences (Grant No. 2014009), the Xie Jialin Foundation of the Institute of High Energy Phsyics, Chinese Academy of Sciences (Grant No. 542019IHEPZZBS10210), the National Basic Research Program (973 Program) of China (Grant No. 2014CB845800), the Swiss Space Office of the State Secretariat for Education, Research and Innovation (ESA PRODEX Programme), and the National Science Center of Poland (Grant No. 2015/17/N/ST9/03556). H.C.L. acknowledge the support from the University of Chinese Academy of Sciences (UCAS) Joint PhD Training Program. M.K. and N.D.A. acknowledge the support of the Swiss National Science Foundation. 

%%%%%%%%%%%%%%%%%%%%%%%%%%%%%%%%%%%%%%%%%%%%%%%%%%

\section*{Data Availability}

The observation data used in this article were obtained from POLAR collaboration (https://www.astro.unige.ch/polar/). The high-level data product derived in this research could be shared upon reasonable request to the corresponding author.



\bibliographystyle{mnras}
\bibliography{main.bib} %  bibtex file is called xxx.bib

%\begin{thebibliography}{99}
%\bibitem[\protect\citeauthoryear{Author}{2013}]{author2013}
%\bibitem[\protect\citeauthoryear{Jones}{2015}]{jones2015}
%Jones C.~D., 2015, Journal of Interesting Stuff, 17, 198
%\bibitem[\protect\citeauthoryear{Smith}{2014}]{smith2014}
%Smith A.~B., 2014, The Example Journal, 12, 345 (Paper I)
%\end{thebibliography}

%%%%%%%%%%%%%%%%%%%%%%%%%%%%%%%%%%%%%%%%%%%%%%%%%%

%%%%%%%%%%%%%%%%% APPENDICES %%%%%%%%%%%%%%%%%%%%%

\appendix
\section{De-rotation method}\label{app:rotation}
	
\begin{figure}
	\centering
	\includegraphics[width=8cm,height=21.0cm]{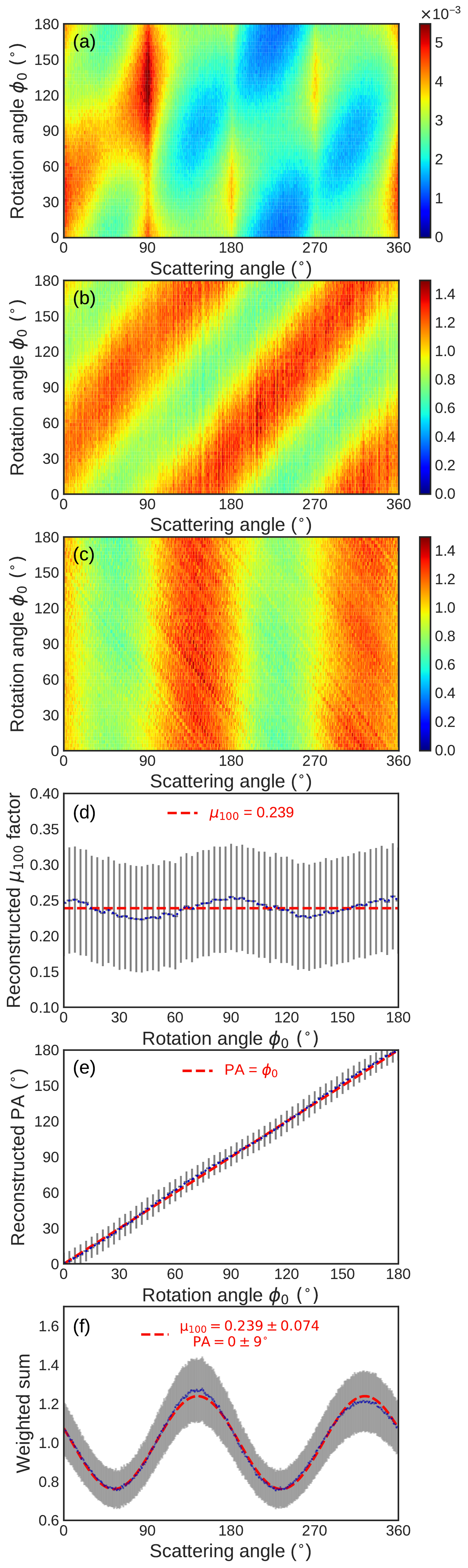}
	\caption{\textbf{The de-rotation method.} (a) is the original (simulated) pattern of the modulation curve v.s. rotation angle $\phi_0$; (b) is the pattern of (a) after geometric-correction; (c) is the pattern after correcting for $\phi_0$ (rotate modulation curves with the corresponding $\phi_0$ values, then $\phi_0=0^{\circ}$) ; (d) is the reconstructed $\mu_{100}$ v.s. $\phi_0$; (e) is the reconstructed PA v.s. $\phi_0$; (f) is the weighted sum of (c) over $\phi_0$ with exposure time as weight.}
	\label{fig:eg_rotation} 
\end{figure}

In order to fix the $\phi_0$ parameter defined in Figure \ref{fig:projection_plane}, a de-rotation process needs to be implemented. Here we show a toy simulation that demonstrates this method. We take a certain incident angle of the source with respect to the detector zenith axis ($\theta=42.6^{\circ}, \phi=52.3^{\circ}$). Then, we  assume the true PD is 100\%, and the $\rm PA_{J}$ is $0^{\circ}$. By dividing $\phi_0$ in bins from $0-180^{\circ}$ with a size of $3^{\circ}$, we obtain 61 bins of $\phi_0$. For each $\phi_0$ bin, the $\phi_0$ value will be directly added into $\rm PA_{J}$, which results in 61 bins of $\rm PA_{S}$. The remaining steps are as follows:

\begin{itemize}
	\item We simulate a sample of PD=100\% \& $\rm PA_{S} = 0:3:180^{\circ}$ and PD=0\% \& $\rm PA_{S} = 0^{\circ}$ (for geometric effect correction) with equal exposure time. Thereafter, we calculate the scattering angles and generate the modulation curves for the sample. The original modulation curves of 100\% PD as a function of $\phi_0$ can be seen in Figure \ref{fig:eg_rotation}(a), where the pattern is dominated by geometric effect and there is no sign of $\phi_0$ dependence;
	
	\item Based on equation (\ref{equ:pol_unp}), we can use the unpolarized modulation curve to correct the geometric effect from Figure \ref{fig:eg_rotation}(a). Note here that the unpolarized curve is independent on $\phi_0$. The geometric-corrected pattern is shown in Figure \ref{fig:eg_rotation}(b). From this pattern, one can clearly see a linear correlation between the modulation curves and $\phi_0$ bins. We further fit the modulation curves along $\phi_0$ bins by using equation (\ref{equ:counts_modu}) and (\ref{equ:mu_factor}). The reconstructed $\mu_{100}$ factor as a function of $\phi_0$ can be found in Figure \ref{fig:eg_rotation}(d), there is small sinusoidal variation (this is more significant for large off-axis incidence which is investigated in Appendix \ref{app:systematic}), but it can be fitted by $\mu_{100}=0.239$. The reconstructed PA ($\rm PA_{S}$) as a function of $\phi_0$ can be found in Figure \ref{fig:eg_rotation}(e), where the PA values are proportional to $\phi_0$;
	
	\item By shifting the modulation curves with the corresponding $\phi_0$ values, we obtain a perfectly aligned pattern which can be found in Figure \ref{fig:eg_rotation}(c). In this pattern, all modulation curves are aligned to $\phi_0=0^{\circ}$, and thus the modulation curves can be stacked. We use the exposure time ratio as a weight to sum over $\phi_0$ bins, the weighted sum of modulation curves is shown in Figure \ref{fig:eg_rotation}(f). By fitting it, we get $\mu_{100}=0.239\pm0.074$ which agrees with Figure \ref{fig:eg_rotation}(d), and $\rm PA=0\pm9 ^{\circ}$ which agrees with assumption of $\rm PA_{J}=0^{\circ}$.
	
\end{itemize}

This toy simulation shows a way to de-rotate $\phi_0$ (makes it fixed to be zero), which is to shift the modulation curves (after geometric effect correction) by the amount corresponding to the $\phi_0$ values. Then the modulation curves can be added together with exposure time ratios as weight. More importantly, the fitting shows that the weighted sum will precisely retain the input polarization information unchanged, which allows us to apply the de-rotation process of $\phi_0$ in our dataset and move forward to further polarization analysis.
    
\section{Systematic effect}\label{app:systematic}

In Appendix \ref{app:rotation}, Figure \ref{fig:eg_rotation}(d) and (e), we see some hints of systematic effects. So before we apply the de-rotation method to our datasets, we further checked the validity of this method for several incident angles, and examined if there are significant systematic effects for the polarization reconstruction. We run a set of toy simulations equal to the one in Appendix \ref{app:rotation} for different incident angles of $\theta$ (10:10:90$^{\circ}$). 
	
\begin{figure}
	\centering
	\includegraphics[width=7.5cm,height=6.5cm]{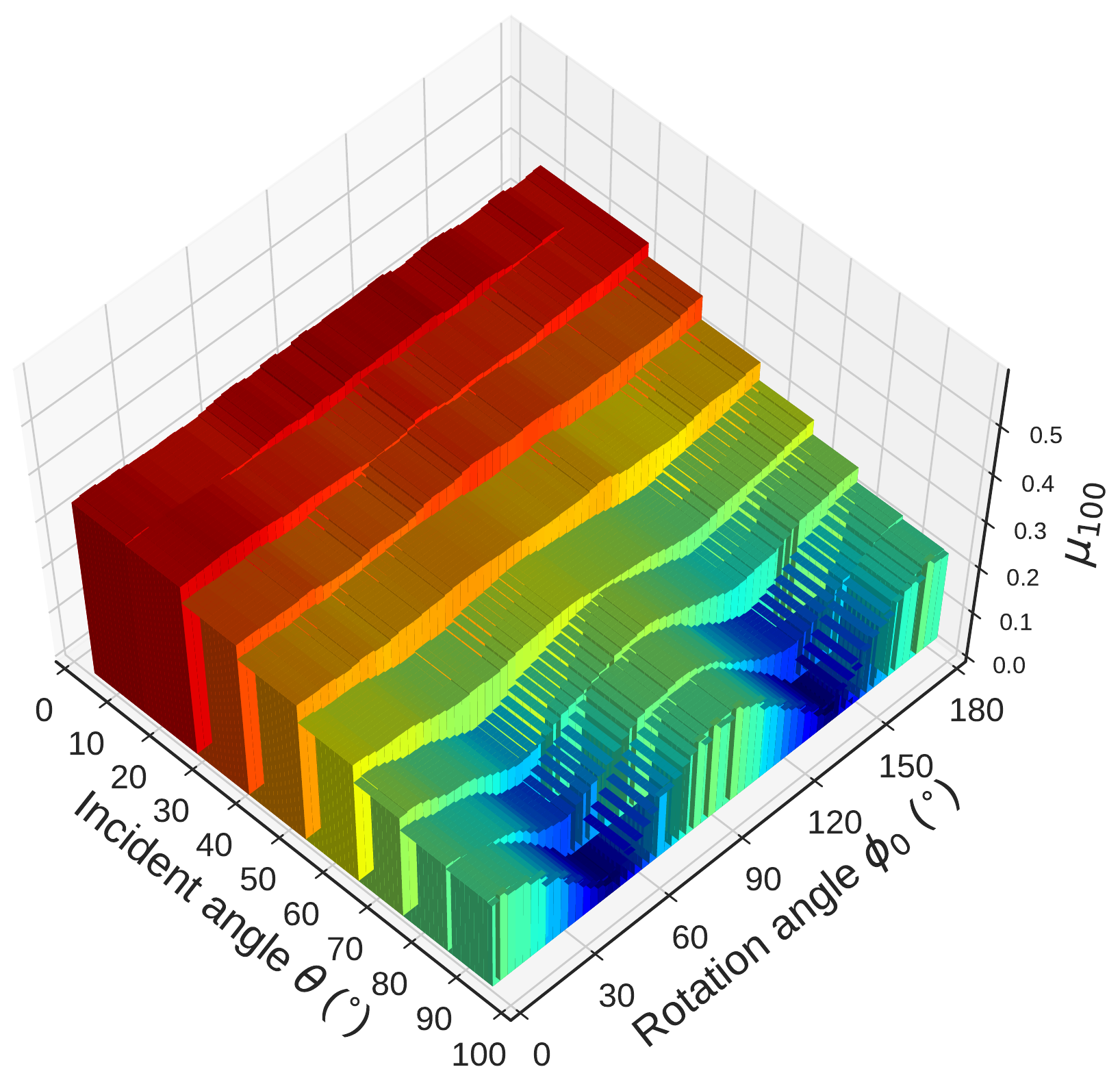}	
	\caption{\textbf{The reconstructed $\mu_{100}$ v.s. rotation angle $\phi_0$ and incident angle $\theta$.} The $\mu_{100}$ factor can be seen to rapidly decrease with $\theta$ value increases especially at some specific $\phi_0$ angles. Additionally, it can be observed to have significant harmonics (additional sinusoidal systematic effect) when $\theta$ is larger.}
	\label{fig:mu100_rotation} 
\end{figure}

\begin{figure}
	\centering
	\includegraphics[width=7.5cm,height=6.5cm]{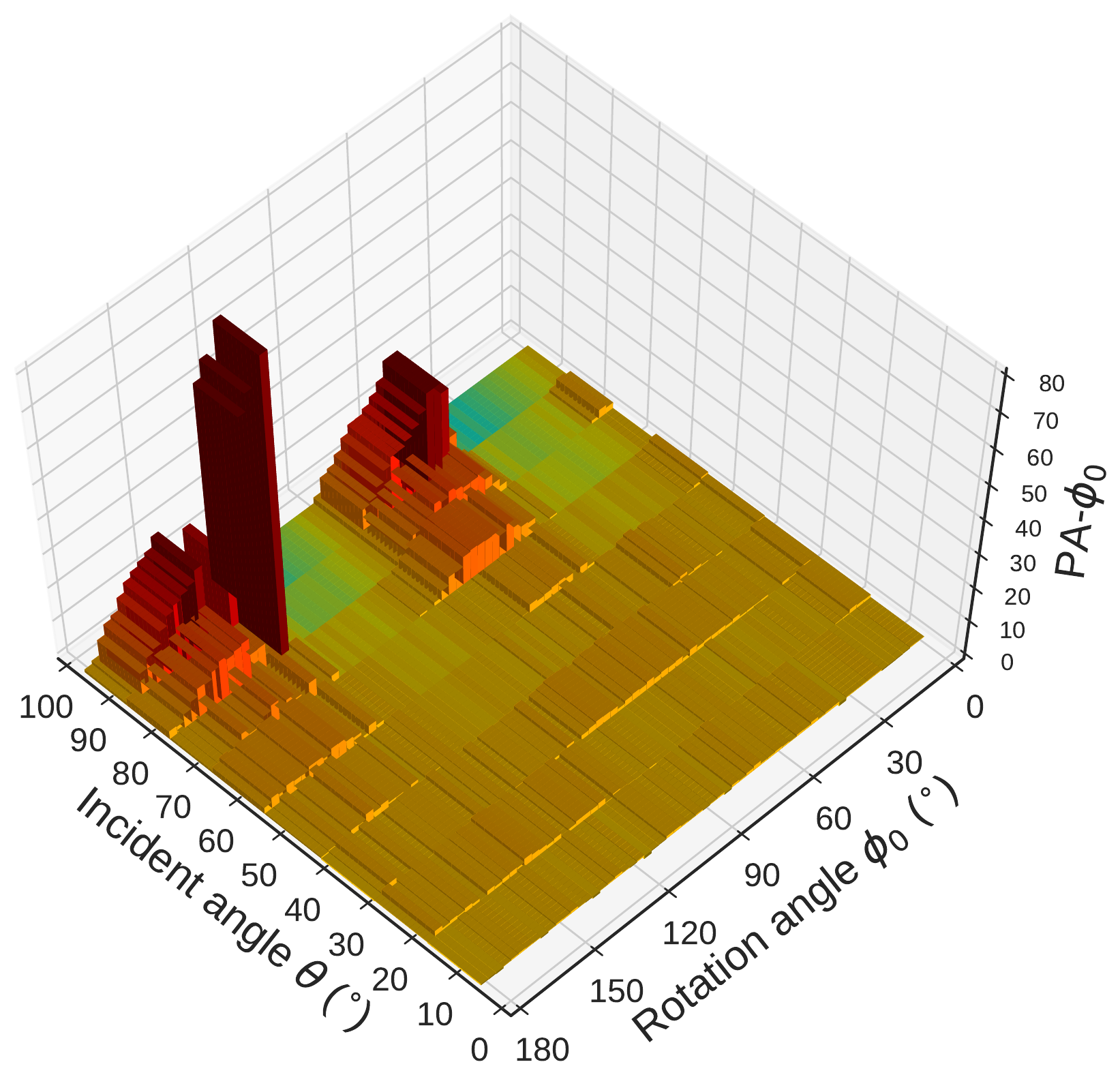}	
	\caption{\textbf{The reconstructed PA minus $\phi_0$ v.s. rotation angle $\phi_0$ and incident angle $\theta$.} The PA has non-negligible effects (additional sinusoidal systematic effect) when $\theta$ is larger than $70^{\circ}$.}
	\label{fig:pa_rotation} 
\end{figure}

The reconstructed $\mu_{100}$ factor as a function of incident angle $\theta$ and rotation angle $\phi_0$ is shown in Figure \ref{fig:mu100_rotation}. There are two pieces of information which can be retrieved from this plot: 1) the $\mu_{100}$ factor rapidly decreases with $\theta$ value; 2) the $\mu_{100}$ factor reconstruction has more significant effects on top of the $180^\circ$ modulation (additional sinusoidal systematic effect) when $\theta$  increases, and when $\theta>70^{\circ}$ this additional effect becomes non-negligible. Based on Figure \ref{fig:eg_rotation} (e), we further define the residual of PA reconstruction, which is PA-$\phi_0$. Figure \ref{fig:pa_rotation} shows the residual as a function of incident angle $\theta$ and rotation angle $\phi_0$. Likewise, we see a similar sinusoidal systematic effect for increasing $\theta$. A stable $\mu_{100}$ factor and good linear PA \& $\phi_0$ correlation are crucial for the de-rotation process and polarization reconstruction. It should be noted here that the additional effects seen in the curve are understood and are a result of photons scattering off materials surrounding the detector. The kinematics of this scattering are dependent on the intrinsic polarization of the photons and therefore remain after correcting with the unpolarized modulation curve. The effect is described in more detail in \citep{Kole2022}. In the work presented here, in order to prevent unwanted systematic effects from potentially entering the analysis we decided to get rid of data when $\theta > 70^{\circ}$. 
	
\section{Direct fitting results}\label{app:chi2}
	
\begin{figure}
	\centering
	\includegraphics[width=8cm,height=15cm]{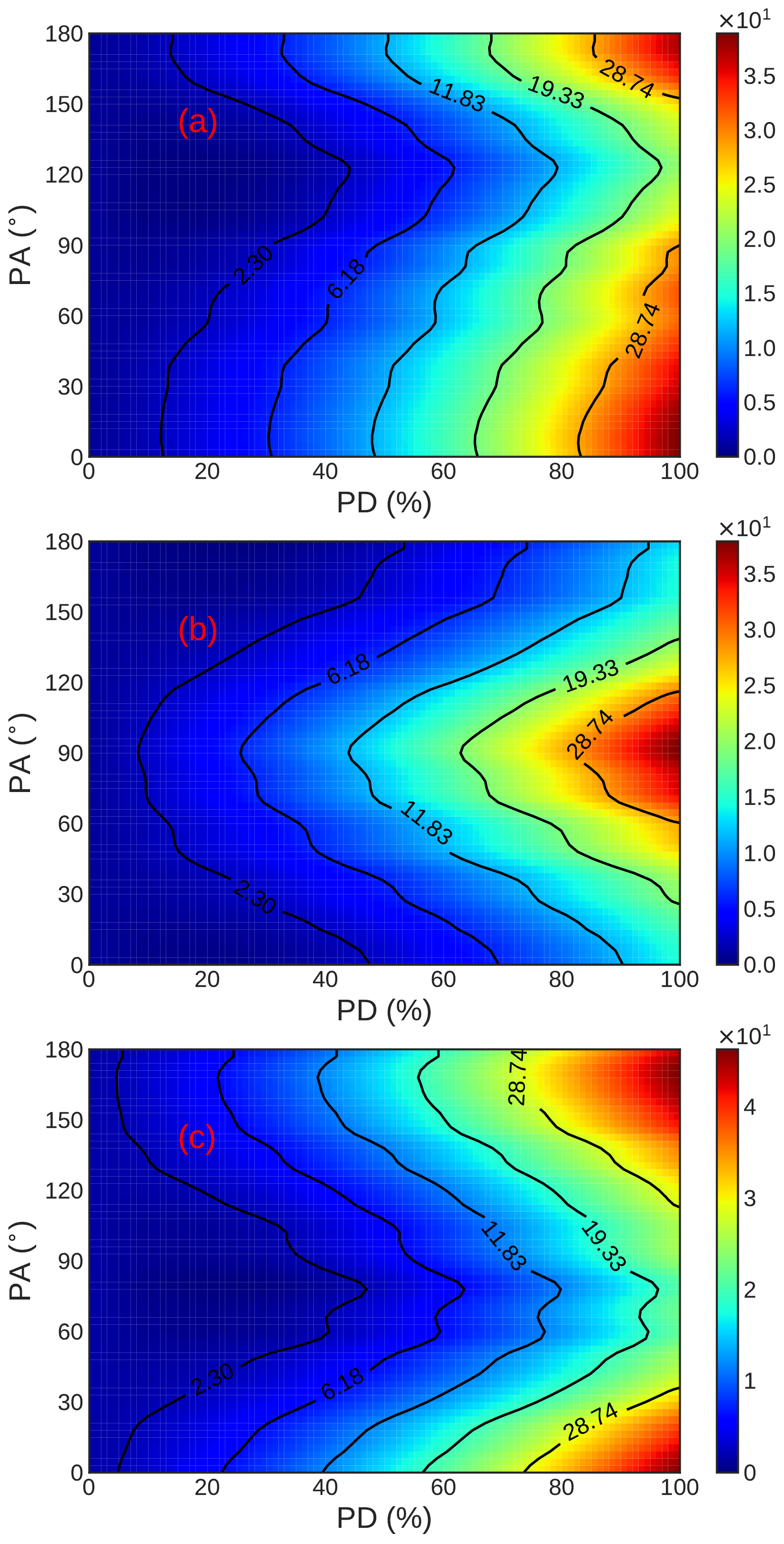}
	\caption{\textbf{The PD/PA $\Delta\chi^2$ contour maps.} Panel (a), (b) and (c) are of the AP, P1 and P2 phase interval respectively. The contour lines in these plots from left to right are $\chi^2$ (degree of freedom is 2) values of 2.3, 6.18, 11.83, 19.33 and 28.74, representing 1-5 sigma confidence levels respectively.}
	\label{fig:pdpa_chi2}
\end{figure}

The joint-fitting among simulated \textbf{B}$_i$ ($\rm PD_i, PA_i$) and observed SRC$'$ generates a $\Delta\chi^2$ contour map which can be used to determine the best fitted value \textbf{A}$_i$ (P\^D, P\^A). The $\Delta\chi^2$ contour maps of AP, P1 and P2 are shown in Figure \ref{fig:pdpa_chi2} (a), (b) and (c) respectively. The contour lines in these plots from left to right are $\chi^2$ (degree of freedom is 2) values of 2.3, 6.18, 11.83, 19.33 and 28.74, representing 1-5 sigma confidence levels respectively.

\section{Tables}\label{app:tables}
	
Here listed two tables including the observation subsample (Table \ref{table:observationODs_papd}) defined by Section \ref{sec:observation} and polarization results (Table \ref{table:PDPA_resultstable}) obtained from this work and the previous works that introduced in Section \ref{sec:introduction}.
	
\onecolumn
\begin{table}
	\footnotesize    %\normalsize, \small, \footnotesize, \scriptsize, \tiny
	\centering
	\caption{\textbf{Subsample of the Crab pulsar observations for polarization analysis}. It is sorted by pulse significance in descending order, and contains 108 HP bins and $\sim$1710 ks exposure times in total.}\label{table:observationODs_papd}
	\newcommand{\tabincell}[2]{\begin{tabular}{@{}#1@{}}#2\end{tabular}}
	\begin{tabular}{ccccc|ccccc}
		\hline
		\tabincell{c}{HP\\No.}	&	\tabincell{c}{$\theta$\\($^{\circ}$)}	&	\tabincell{c}{$\phi$\\($^{\circ}$)}	&	\tabincell{c}{significance\\($\sigma$)}	&	\tabincell{c}{exposure time\\(s)}	& \tabincell{c}{HP\\No.}	&	\tabincell{c}{$\theta$\\($^{\circ}$)}	&	\tabincell{c}{$\phi$\\($^{\circ}$)}	&	\tabincell{c}{significance\\($\sigma$)}	&	\tabincell{c}{exposure time\\(s)} \\ 
		\hline
		190	&	60.00	&	163.13	&	5.37	&	23030	&	88	&	41.86	&	57.86	&	2.99	&	13640	\\
		213	&	65.38	&	56.25	&	5.32	&	22109	&	89	&	41.86	&	70.71	&	2.98	&	14883	\\
		102	&	41.86	&	237.86	&	5.16	&	9090	&	117	&	48.19	&	61.88	&	2.92	&	14919	\\
		105	&	41.86	&	276.43	&	4.92	&	50118	&	73	&	35.66	&	202.50	&	2.92	&	8284	\\
		83	&	35.66	&	352.50	&	4.88	&	26669	&	65	&	35.66	&	82.50	&	2.92	&	8041	\\
		150	&	54.31	&	67.50	&	4.79	&	19040	&	165	&	54.31	&	236.25	&	2.85	&	8155	\\
		221	&	65.38	&	146.25	&	4.75	&	12177	&	215	&	65.38	&	78.75	&	2.82	&	11400	\\
		148	&	54.31	&	45.00	&	4.73	&	24194	&	156	&	54.31	&	135.00	&	2.82	&	9319	\\
		5	&	11.72	&	67.50	&	4.38	&	12591	&	157	&	54.31	&	146.25	&	2.74	&	9928	\\
		92	&	41.86	&	109.29	&	4.24	&	32009	&	93	&	41.86	&	122.14	&	2.74	&	12605	\\
		78	&	35.66	&	277.50	&	4.21	&	36787	&	184	&	60.00	&	95.63	&	2.73	&	16470	\\
		168	&	54.31	&	270.00	&	4.20	&	27964	&	24	&	23.56	&	11.25	&	2.71	&	17758	\\
		111	&	41.86	&	353.57	&	4.16	&	23033	&	119	&	48.19	&	84.38	&	2.71	&	11207	\\
		180	&	60.00	&	50.63	&	4.15	&	17433	&	199	&	60.00	&	264.38	&	2.70	&	18473	\\
		182	&	60.00	&	73.13	&	4.01	&	17415	&	4	&	11.72	&	22.50	&	2.70	&	16194	\\
		40	&	29.57	&	9.00	&	4.00	&	23905	&	19	&	17.61	&	225.00	&	2.70	&	9850	\\
		6	&	11.72	&	112.50	&	3.99	&	10384	&	10	&	11.72	&	292.50	&	2.69	&	12292	\\
		66	&	35.66	&	97.50	&	3.98	&	10954	&	134	&	48.19	&	253.13	&	2.67	&	20281	\\
		160	&	54.31	&	180.00	&	3.94	&	11043	&	12	&	17.61	&	15.00	&	2.67	&	16109	\\
		18	&	17.61	&	195.00	&	3.87	&	9885	&	48	&	29.57	&	153.00	&	2.65	&	9711	\\
		181	&	60.00	&	61.88	&	3.79	&	19086	&	127	&	48.19	&	174.38	&	2.65	&	7039	\\
		125	&	48.19	&	151.88	&	3.74	&	17653	&	217	&	65.38	&	101.25	&	2.64	&	18412	\\
		68	&	35.66	&	127.50	&	3.73	&	9752	&	22	&	17.61	&	315.00	&	2.58	&	16647	\\
		23	&	17.61	&	345.00	&	3.73	&	15819	&	219	&	65.38	&	123.75	&	2.58	&	6078	\\
		232	&	65.38	&	270.00	&	3.72	&	25103	&	137	&	48.19	&	286.88	&	2.57	&	41851	\\
		62	&	35.66	&	37.50	&	3.71	&	13127	&	16	&	17.61	&	135.00	&	2.56	&	9402	\\
		11	&	11.72	&	337.50	&	3.67	&	15539	&	149	&	54.31	&	56.25	&	2.56	&	18399	\\
		28	&	23.56	&	101.25	&	3.66	&	9133	&	86	&	41.86	&	32.14	&	2.52	&	19243	\\
		212	&	65.38	&	45.00	&	3.63	&	26595	&	59	&	29.57	&	351.00	&	2.52	&	28498	\\
		116	&	48.19	&	50.63	&	3.53	&	15189	&	216	&	65.38	&	90.00	&	2.42	&	11091	\\
		106	&	41.86	&	289.29	&	3.53	&	30453	&	29	&	23.56	&	123.75	&	2.41	&	8225	\\
		67	&	35.66	&	112.50	&	3.52	&	10780	&	214	&	65.38	&	67.50	&	2.39	&	18448	\\
		61	&	35.66	&	22.50	&	3.51	&	21940	&	44	&	29.57	&	81.00	&	2.36	&	8741	\\
		27	&	23.56	&	78.75	&	3.50	&	10359	&	90	&	41.86	&	83.57	&	2.33	&	11715	\\
		122	&	48.19	&	118.13	&	3.48	&	22075	&	124	&	48.19	&	140.63	&	2.32	&	13179	\\
		77	&	35.66	&	262.50	&	3.46	&	23034	&	17	&	17.61	&	165.00	&	2.30	&	9994	\\
		0	&	5.85	&	45.00	&	3.45	&	11652	&	87	&	41.86	&	45.00	&	2.27	&	13121	\\
		30	&	23.56	&	146.25	&	3.44	&	7739	&	33	&	23.56	&	213.75	&	2.25	&	8592	\\
		42	&	29.57	&	45.00	&	3.44	&	14567	&	14	&	17.61	&	75.00	&	2.23	&	11740	\\
		115	&	48.19	&	39.38	&	3.42	&	21977	&	151	&	54.31	&	78.75	&	2.17	&	12437	\\
		169	&	54.31	&	281.25	&	3.40	&	17353	&	104	&	41.86	&	263.57	&	2.16	&	25649	\\
		152	&	54.31	&	90.00	&	3.36	&	11146	&	81	&	35.66	&	322.50	&	2.14	&	13400	\\
	    185	&	60.00	&	106.88	&	3.31	&	18461	&	135	&	48.19	&	264.38	&	2.11	&	21682	\\
		15	&	17.61	&	105.00	&	3.24	&	9896	&	8	&	11.72	&	202.50	&	2.10	&	11479	\\
		20	&	17.61	&	255.00	&	3.22	&	9674	&	193	&	60.00	&	196.88	&	2.09	&	3006	\\
		43	&	29.57	&	63.00	&	3.22	&	13808	&	123	&	48.19	&	129.38	&	2.08	&	11249	\\
		26	&	23.56	&	56.25	&	3.16	&	15508	&	63	&	35.66	&	52.50	&	2.07	&	13838	\\
		187	&	60.00	&	129.38	&	3.13	&	6616	&	41	&	29.57	&	27.00	&	2.06	&	14790	\\
		121	&	48.19	&	106.88	&	3.13	&	20974	&	120	&	48.19	&	95.63	&	2.06	&	15626	\\
		71	&	35.66	&	172.50	&	3.10	&	6685	&	158	&	54.31	&	157.50	&	2.04	&	23707	\\
		69	&	35.66	&	142.50	&	3.06	&	10878	&	166	&	54.31	&	247.50	&	2.03	&	8921	\\
		7	&	11.72	&	157.50	&	3.05	&	11112	&	21	&	17.61	&	285.00	&	2.02	&	12400	\\
		36	&	23.56	&	281.25	&	3.05	&	12717	&	161	&	54.31	&	191.25	&	2.01	&	10340	\\
		136	&	48.19	&	275.63	&	3.01	&	33379	&	189	&	60.00	&	151.88	&	2.01	&	12645	\\
		\hline
	\end{tabular}
\end{table}
	
\begin{landscape}
	%\begin{sidewaystable*}
	\begin{table}
	\footnotesize      %\normalsize, \small, \footnotesize, \scriptsize, \tiny
	\centering
	\caption{\textbf{Polarization measurements of the Crab pulsar \& nebula by POLAR and other instruments}}\label{table:PDPA_resultstable}
	\newcommand{\tabincell}[2]{\begin{tabular}{@{}#1@{}}#2\end{tabular}}
	\begin{tabular}{|c|c|c|c|c|c|c|c|c|}
		\hline
		Mission	&	Publication	&	\tabincell{c}{Energy band\\(keV)}	&	\tabincell{c}{Exposure time\\(ks)}	& Phase range	&	PA ($^\circ$)	&	PD (\%)	& \tabincell{c}{PD$_{\rm down}$\%\\(confidence\%)} &	\tabincell{c}{PD$_{\rm up}$\%\\(confidence\%)} \\ 
		\hline
		\multirow{3}{*}{POLAR}	&	\multirow{3}{*}{this work}	&	\multirow{3}{*}{50--500}	&	\multirow{3}{*}{$\sim$1710}	&	\tabincell{c}{AP (-n)}	&	$108^{+33}_{-54}$	&	$14^{+15}_{-10}$	&	\multirow{3}{*}{N$/$A}	&	55 (99.7)	\\
		\cline{5-7}\cline{9-9}
		&		&	&		&	P1	&	$174^{+39}_{-36}$	&	$17^{+18}_{-12}$	&	&	66 (99.7)	\\
		\cline{5-7}\cline{9-9}
		&		&	&		&	P2	&	$78^{+39}_{-30}$	&	$16^{+16}_{-11}$	&	&	57 (99.7)	\\
		\hline
		\multirow{3}{*}{PolarLight}	&	\multirow{3}{*}{\cite{2020NatAs...4..511F}}	&	\multirow{3}{*}{3.0--4.5}	&	\multirow{3}{*}{$\sim$660}	&	\tabincell{c}{AP (+n)}	&	145.8 $\pm$ 5.7	&	$15.3^{+3.1}_{-3.0}$	&	\multirow{3}{*}{N$/$A}	&	\multirow{3}{*}{N$/$A}	\\
		\cline{5-7}
		&		&	&		&	P1+P2	&	147.6 $\pm$ 7.0	&	15.8 $\pm$ 3.9	&	&	\\
		\cline{5-7}
		&		&	&		&	OP	&	142.4 $\pm$ 11.0	&	$14.0^{+5.2}_{-5.4}$	&	&	\\
		\hline
		
		\tabincell{c}{Hitomi\\SGD}	&	\cite{2018PASJ...70..113H}	&	 60--160	&	5	&	\tabincell{c}{AP (+n)}	&	$110.7^{+13.2}_{-13.0}$	&	22.1 $\pm$ 10.6	&	N$/$A	&	N$/$A	\\
		\hline
		\multirow{2}{*}{\tabincell{c}{Astrosat\\CZTI}}	&	\multirow{2}{*}{\cite{2018NatAs...2...50V}}	&	\multirow{2}{*}{100--380}	&	\multirow{2}{*}{800}	&	\tabincell{c}{AP (+n)}	&	143.5 $\pm$ 2.8	&	32.7 $\pm$ 5.8	&	\multirow{2}{*}{\tabincell{c}{N$/$A}}	&	\multirow{2}{*}{\tabincell{c}{N$/$A}}	\\
		\cline{5-7}
		%&		&	&		&	P1	&	159.52 $\pm$ 1.3 $^{*}$	&	15.386 $\pm$ 2.1 $^{*}$	&	&	\\
		%\cline{5-7}
		%&		&	&		&	P2	&	144.293 $\pm$ 1.1 $^{*}$	&	21.723 $\pm$ 2.2 $^{*}$	&	&	\\
		%\cline{5-9}
		&		&	&		&	OP	&	140.9 $\pm$ 3.7	&	39.0 $\pm$ 10.0	&	&	\\
		\hline
		\multirow{4}{*}{PoGO+}	&	\multirow{4}{*}{\cite{2017NatSR...7.7816C}}	&	\multirow{4}{*}{20--160}	&	\multirow{4}{*}{92}	&	\tabincell{c}{AP (+n)}	&	131.3 $\pm$ 6.8	&	20.9 $\pm$ 5.0	&	\multirow{4}{*}{\tabincell{c}{N$/$A}}	&	32.5(99)	\\
		\cline{5-7}\cline{9-9}
		&		&	&		&	P1	&	153 $\pm$ 43	&	$0^{+29}_{-0}$ (20.0$\pm$24.8)	&	&	72.6 (99)	\\
		\cline{5-7}\cline{9-9}
		&		&	&		&	P2	&	86 $\pm$ 18	&	$33.5^{+18.6}_{-22.3}$	&	&	80.9 (99)	\\
		\cline{5-7}\cline{9-9}
		&		&	&		&	OP	&	137 $\pm$ 15	&	$17.4^{+8.6}_{-9.3}$	&	&	37.3 (99)	\\
		\hline
		\multirow{3}{*}{\tabincell{c}{INTEGRAL\\SPI}} & \cite{2019ApJ...882..129J} &  130--400 & $\sim$6400 & \tabincell{c}{AP (+n)} & 120 $\pm$ 6 & 24 $\pm$ 4 & \multirow{3}{*}{N$/$A} & \multirow{3}{*}{N$/$A} \\
		\cline{2-7}
		& \cite{2013ApJ...769..137C} &  130--400 & 600 & AP (+n) & 117 $\pm$ 9 & 28 $\pm$ 6 & &  \\
		\cline{2-7}
		&	\cite{2008Sci...321.1183D}	&	 100--1000	&	N$/$A	&	OP	&	123 $\pm$ 11	&	47 $\pm$ 10	&	&	\\
		\hline
		\multirow{3}{*}{\tabincell{c}{INTEGRAL\\IBIS}}	&	\multirow{3}{*}{\cite{2008ApJ...688L..29F}}	&	\multirow{3}{*}{200--800}	&	\multirow{3}{*}{1200}	&	\tabincell{c}{AP (+n)}	&	100 $\pm$ 11	&	$47^{+19}_{-13}$	&	\multirow{2}{*}{N$/$A}	&	\multirow{3}{*}{N$/$A}	\\
		\cline{5-7}
		&		&	&		&	P1+P2	&	70 $\pm$ 20	&	$42^{+30}_{-16}$	&	&		\\
		\cline{5-8}
		&		&	&		&	OP	&	120.6 $\pm$ 8.5	&	N$/$A	&	72 (95)	&		\\
		\hline
		\multirow{4}{*}{\tabincell{c}{OSO}}	&	\multirow{2}{*}{\cite{1978ApJ...220L.117W}}	&	2.6	&	\multirow{2}{*}{$\sim$256}	&	\multirow{2}{*}{\tabincell{c}{OP}}	&	156.4 $\pm$ 1.4	&	19.2 $\pm$ 1.0	&	\multirow{4}{*}{N$/$A}	&	\multirow{4}{*}{N$/$A}	\\
		\cline{3-3}\cline{6-7}
		&		&	5.2	&		&	&	152.6 $\pm$ 4.0	&	19.5 $\pm$ 2.8	&	&		\\
		\cline{2-7}
		&	\multirow{2}{*}{\cite{1976ApJ...208L.125W}}	&	2.6	&	\multirow{2}{*}{N$/$A}	&	\multirow{2}{*}{\tabincell{c}{AP (+n)}}	&	161.1 $\pm$ 2.8	&	15.7 $\pm$ 1.5	&	&	\\
		\cline{3-3}\cline{6-7}
		&		&	5.2	&		&	&	155.5 $\pm$ 6.6	&	18.3 $\pm$ 4.2	&	&		\\
		\hline
		\multicolumn{9}{|l|}{\tabincell{l}{Notes:\\ 1. $^{+\uparrow}_{-\downarrow}$ or $\pm\updownarrow$: one sigma up(+)/down(-) deviation; \\ 2. PD$_{\rm down}$ (99\%) \& PD$_{\rm up}$ (99\%): upper and lower limit for PD with 99\% confident level respectively; \\ 3. $\pm$n: the Crab nebula was(+) included in Averaged-Phase (AP) or not(-); \\ 4. OP: off pulse, represent for the Crab nebula; \\5. N$/$A: not been shown from the publication; \\6. The PD of P1 by PoGO+ is $0^{+29}_{-0}$\% with the marginalized estimate and  20.0$\pm$24.8\% (shown in Figure \ref{fig:pa_onto_crab}) with the maximum a posterior probability estimate.} }\\
		\hline
	\end{tabular}
	\end{table} 
	%\end{sidewaystable*}
%%%%%%%%%%%%%%%%%%%%%%%%%%%%%%%%%%%%%%%%%%%%%%%%%%
    % Don't change these lines
    \bsp	% typesetting comment
    \label{lastpage}
\end{landscape}

\end{document}